\documentclass[11pt,psfig,epsf,bbox]{article} 

\usepackage{amsmath,amssymb,amsthm}
\DeclareMathAlphabet{\mathpzc}{OT1}{pzc}{m}{it}

\begin{document}

\newcommand{\vAi}{{\cal A}_{i_1\cdots i_n}} \newcommand{\vAim}{{\cal
A}_{i_1\cdots i_{n-1}}} \newcommand{\vAbi}{\bar{\cal A}^{i_1\cdots i_n}}
\newcommand{\vAbim}{\bar{\cal A}^{i_1\cdots i_{n-1}}}
\newcommand{\htS}{\hat{S}} \newcommand{\htR}{\hat{R}}
\newcommand{\htB}{\hat{B}} \newcommand{\htD}{\hat{D}}
\newcommand{\htV}{\hat{V}} \newcommand{\cT}{{\cal T}} \newcommand{\cM}{{\cal
M}} \newcommand{\cMs}{{\cal M}^*}
 \newcommand{\vk}{{\bf k}}
\newcommand{\vK}{{\vec K}} \newcommand{\vb}{{\vec b}} \newcommand{{\vp}}{{\vec
p}} \newcommand{{\vq}}{{\vec q}} \newcommand{\vQ}{{\vec Q}}
\newcommand{\vx}{{\vec x}}
\newcommand{\tr}{{{\rm Tr}}} 
\newcommand{\beq}{\begin{equation}}
\newcommand{\eeq}[1]{\label{#1} \end{equation}} 
\newcommand{\half}{{\textstyle
\frac{1}{2}}} \newcommand{\gton}{\stackrel{>}{\sim}}
\newcommand{\lton}{\mathrel{\lower.9ex \hbox{$\stackrel{\displaystyle
<}{\sim}$}}} \newcommand{\ee}{\end{equation}}
\newcommand{\ben}{\begin{enumerate}} \newcommand{\een}{\end{enumerate}}
\newcommand{\bit}{\begin{itemize}} \newcommand{\eit}{\end{itemize}}
\newcommand{\bc}{\begin{center}} \newcommand{\ec}{\end{center}}
\newcommand{\bea}{\begin{eqnarray}} \newcommand{\eea}{\end{eqnarray}}
\newcommand{\beqar}{\begin{eqnarray}} \newcommand{\eeqar}[1]{\label{#1}
\end{eqnarray}} \newcommand{\bra}[1]{\langle {#1}|}
\newcommand{\ket}[1]{|{#1}\rangle}
\newcommand{\norm}[2]{\langle{#1}|{#2}\rangle}
\newcommand{\brac}[3]{\langle{#1}|{#2}|{#3}\rangle} \newcommand{\hilb}{{\cal
H}} \newcommand{\pleft}{\stackrel{\leftarrow}{\partial}}
\newcommand{\pright}{\stackrel{\rightarrow}{\partial}}

\begin{center}
{\Large {\bf{Collisional Energy Loss in a Finite Size QCD Matter}}}

\vspace{1cm}

{ Magdalena Djordjevic}

\vspace{.8cm}

{\em { Dept. Physics, Ohio State University, 191 West Woodruff Avenue, 
Columbus, OH 43210, USA}}

\vspace{.5cm}

\today
\end{center}

\vspace{.5cm}

\begin{abstract}
Computation of collisional energy loss in a finite size QCD medium has become 
crucial to obtain reliable predictions for jet quenching in ultra-relativistic 
heavy ion collisions. We here compute this energy loss up to the zeroth order 
in opacity. Our approach consistently treats both soft and hard contributions 
to the collisional energy loss. Consequently, it removes the unphysical energy 
gain in a region of lower momenta obtained by previous computations. Most 
importantly, we show that for characteristic QCD medium scales, finite 
size effects on the collisional energy loss are not significant.
\end{abstract}

\section{Introduction}

The suppression pattern of high transverse momentum hadrons is a powerful tool 
to map out the density of a QCD plasma created in ultra-relativistic heavy ion 
collisions (URHIC)~\cite{Gyulassy_2002}-\cite{Gyulassy:1991xb}. This 
suppression (called jet quenching) is considered to be mainly a consequence of 
medium induced radiative energy loss of high energy 
partons~\cite{MVWZ:2004}-\cite{KW:2004}. However, recent non-photonic 
single electron data~\cite{Adler:2005xv,elecQM05_STAR} (which present an 
indirect probe of heavy quark energy loss) showed large disagreement with 
the radiative energy loss predictions, as long as realistic values of 
parameters are assumed~\cite{Djordjevic:2005db}. This raised the question 
of what is the cause for the observed discrepancy.

Recent studies~\cite{Mustafa,Dutt-Mazumder} suggested that one of the basic 
assumptions that pQCD collisional energy loss is negligible compared to 
radiative~\cite{Bjorken:1982tu} may be incorrect. 
In~\cite{Mustafa,Dutt-Mazumder} it was shown that, for a range of parameters 
relevant for RHIC, radiative and collisional energy losses for heavy quarks 
were in fact comparable to each other, and therefore collisional energy loss 
can not be neglected in the computation of jet quenching. This result came as 
a surprise because from the earlier 
estimates~\cite{Bjorken:1982tu}-~\cite{Lin:1997cn}, the typical collisional 
energy loss was erroneously considered to be small compared to the radiative 
energy loss. In~\cite{WHDG} it was consequently suggested that the inclusion 
of collisional energy loss may help in solving the ``single electron puzzle''. 
However, in that study (as well as~\cite{Mustafa}-\cite{Lin:1997cn}) finite 
size effects were not taken into account. 

A recent paper by Peigne {\em et al.}~\cite{Peigne} is the first study that 
made an attempt to include finite size effects in the collisional energy 
loss. This work suggested that collisional energy loss is large only in the 
ideal infinite medium case, while finite size effects lead to a significant 
reduction of the collisional energy loss. However, this paper did not 
completely separate collisional and radiative energy loss effects. 
Consequently, it remained unclear how important are the finite size effects on 
the collisional energy loss. 

Therefore, it became necessary to consistently compute (only) the collisional 
energy loss in finite size QCD medium. Additionally, this paper aims to 
address whether -and to what extent- there is an over-counting between 
collisional and radiative energy loss computations. 

The outline of the paper is as follows: In Section~2, we will compute the 
collisional energy loss in a finite size QCD medium. In Section~3, we will 
consider the special case when a particle is produced at $x_0=-\infty$ 
(infinite medium case). We will show that in special limits, our calculations 
recover previous results~\cite{TG,BT}. However, contrary to~\cite{TG,BT}, our 
computation does not encounter unphysical energy gain in the low momentum 
region~\cite{TG,BT}. In Section~4 we will give a numerical study of the 
collisional energy loss in both finite and infinite QCD medium. Contrary to the
results obtained in Peigne {\em et al.}~\cite{Peigne} we will show that finite 
size effects do not have a significant effect on the collisional energy loss. 
The conclusions and outlook are given in Section~5.

\section{Collisional energy loss in finite size QCD medium} 

In this Section we will compute the collisional energy loss (up to the 
$0^{th}$ order in opacity) when the jet is produced in a finite size 
dielectric medium. The contribution to this energy loss comes from one 
Hard-Thermal Loop (HTL) gluon propagator (see Appendix A), which is the reason 
why we call it the $0^{th}$ order in opacity collisional energy loss (note the 
analogy with the $0^{th}$ order in opacity radiative energy 
loss~\cite{DG_TM}-\cite{MD_TR}, which is further discussed in Appendix A).

\begin{figure}[h]
\vspace*{3cm} \includegraphics{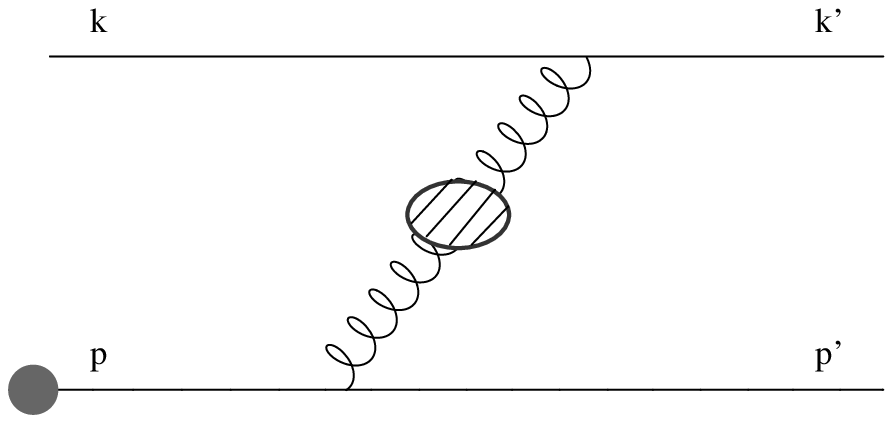}
\caption{Feynman diagram for the amplitude that contributes to the collisional 
energy loss in QCD medium. The large dashed circle (``blob'') represents the 
effective gluon propagator~\cite{DG_TM}.}
\label{Diag_EL}
\end{figure}

In this computation we use the most intuitive approach, i.e. we compute the 
diagram $|M_{el}|$ shown in Fig.~\ref{Diag_EL}. Note that the ``blob'' 
represents the effective gluon propagator. A proof of the validity of this 
approach is given in Appendix A. This approach has already been used without 
proof in~\cite{BT_fermions,Baym}, under the justification that it reproduces 
the same results as the imaginary time formalism.

Similarly as in~\cite{MD_TR} we introduce the finite size medium by starting 
from the approach described in~\cite{Zakharov}. We consider a static medium of 
size $L$, and assume that collisional energy loss can occur 
only inside the medium. The Feynman diagram $|M_{el}|$ (see Fig.~\ref{Diag_EL})
then represents the source $J$, which at time $x_0$ produces an 
off-shell jet with momentum $p$, and subsequently (at 
$ x_1 >x_0$) exchanges a virtual gluon with parton in the 
medium with momentum $k$. The jet and the medium parton emerge with 
momentum $p^\prime$ and $k^\prime$ respectively. Since our focus is on
heavy quark jets with mass $M$, we here neglect the thermal shifts of the 
heavy quark mass.

We assume, as in~\cite{GLV}, that $J$ changes slowly with $p^\prime$, i.e. 
that $J(p^\prime+q)\approx J(p)$. Since we consider both soft and hard 
contributions, we take into account spin effects. The computation that we 
present in this paper is gauge invariant~\cite{BT_fermions}, but for 
simplicity we further use Coulomb gauge.

The effective gluon propagator shown in Fig. 1 has both transverse and 
longitudinal contributions~\cite{Kalashnikov:cy}-\cite{Gyulassy_Selikhov}. In 
Coulomb gauge the gluon propagator has the fallowing form:

\beq
D^{\mu \nu } (\omega, \vec{\mathbf{q}})=
- P^{\mu \nu } \Delta_T (\omega, \vec{\mathbf{q}})-
Q^{\mu \nu } \Delta_{L}(\omega, \vec{\mathbf{q}}),
\eeq{dmnMed}
where $q=(\omega, \vec{\mathbf{q}})$ is the 4-momentum of the gluon, 
while $\Delta_T$ and $\Delta_L$ are effective transverse and longitudinal 
gluon propagators given by~\cite{Gyulassy_Selikhov}:

\beq
\Delta_T^{-1} = \omega^2 - \vec{\mathbf{q}}^{2} - \frac{\mu^{2}}{2} -
\frac{(\omega ^{2} - \vec{\mathbf{q}}^{2})\mu^{2}}{2 \vec{\mathbf{q}}^{2}} 
(1+\frac{\omega }{2|\vec{\mathbf{q}}|}
\ln |\frac{\omega -|\vec{\mathbf{q}}|}{\omega +|\vec{\mathbf{q}}|}|),
\eeq{DeltaT}

\bigskip

\beq
\Delta_{L}^{-1}= \vec{\mathbf{q}}^{2}+ \mu^{2}
(1+\frac{\omega }{2|\vec{\mathbf{q}}|} 
\ln |\frac{\omega -|\vec{\mathbf{q}}|}{\omega +|\vec{\mathbf{q}}|}|),
\eeq{DeltaL}
where $\mu_D^2=g^2 T^2 (1+\frac{N_f}{6})$ is the Debye mass.

The only nonzero terms in transverse $(P_{\mu \nu })$ and longitudinal 
$(Q_{\mu \nu })$ projectors are the following:

\beq
P^{i j}=\delta^{ij} - \frac{q^i  q^j}{|\vec{\mathbf{q}}|^2},
\eeq{Pij}

\beq
Q^{0 0 }=1.
\eeq{Q00}

The matrix element for this $0^{th}$ order in opacity collisional process can 
then be written in the following form (for simplicity we here omit color 
factors, whose contribution we will add in the end)

\beqar
i M_{el} &=&  \int d^4x_0 \; J(x_0) \; d^4  x_1 
\;(-i) \int  \frac{d^3 \vec{\mathbf{p}}}{(2 \pi)^3 2 E} \; \Theta(t_1- t_0)\; 
\Theta(L/v - (t_1- t_0))\; \,\nonumber \\ &&\;
\bar{u} (p^\prime, s^\prime)  e^{i p^\prime x_1}\; i g \gamma^\mu\; 
u (p, s) e^{- i p x_1} 
\; \int d^4 x_2 \; (-i) \int \frac{d^4 q}{(2 \pi)^4} 
D_{\mu \nu } (q) \, e^{-i q (x_2-x_1)} \;\,\nonumber \\ && \bar{u} 
(k^\prime, \lambda^\prime) e^{i k^\prime x_2}\; i g \gamma^\nu \;
u (k, \lambda) e^{-i k x_2} \; .
\eeqar{eq:1}

Here $p$, $s$, $k$ and $\lambda$ are the 4-momenta and spins of the incoming 
jet and medium parton, while the corresponding primed variables correspond to 
outgoing jet and medium parton (the medium parton can be quark, antiquark or 
gluon). The medium partons are considered to be massless, i.e. 4-momentum 
$k$ ($k^\prime$) is assumed to be $k=(|\vec{\mathbf{k}}|,\vec{\mathbf{k}})$ 
($k^\prime=(|\vec{\mathbf{k^\prime}}|,\vec{\mathbf{k^\prime}})$). The 
condition that the interaction between jet and medium parton has to occur 
inside the QCD medium of finite size $L$ is implemented through the second 
$\theta$ function giving maximal interaction time of $(t_1-t_0)_{max}=L/v$. 
We will further define $x \equiv x_1-x_0=(t,\vec{\mathbf{x}})$.

\medskip 

The Eq.~(\ref{eq:1}) simplifies to
\beqar
i M_{el} &=&  g^2 \int  \frac{d^3 \vec{\mathbf{p}}}{(2 \pi)^3 2 E}
\int \frac{d^4 q}{(2 \pi)^4} \int d^4x_0 \; J(x_0) 
e^{i (p^\prime +q) x_0} \int d^3 \vec{\mathbf{x}} \; \int_0^{L/v} d t \;
e^{- i (p- p^\prime -q) x} 
\nonumber \\ &&\; (2 \pi)^4 \delta(k^\prime-k-q)\, 
D_{\mu \nu } (q)\, \bar{u} (p^\prime, s^\prime) \gamma^\mu u (p, s) 
\bar{u} (k^\prime, \lambda^\prime) \gamma^\nu u (k, \lambda)\nonumber \\
&=& J(p^\prime) \; \frac{1}{2 E} 
\frac{1- e^{- i (E- E^\prime -\omega) L/v}}{E- E^\prime -\omega} i \cal{M}.
\eeqar{eq:2} 
where $E=\sqrt{M^2+\vec{\mathbf{p}}^2}$, $M$ is the jet mass,
$\vec{\mathbf{p}}= \vec{\mathbf{p^\prime}}- 
(\vec{\mathbf{k^\prime}}-\vec{\mathbf{k}})$ and 
$\omega=|\vec{\mathbf{k^\prime}}|-|\vec{\mathbf{k}}|$.

In the last step we used $J(p^\prime+q)\approx J(p^\prime)$~\cite{GLV} and 
defined
\beqar
{\cal M}=g^2 
D_{\mu \nu } (k^\prime-k) \bar{u} (p^\prime, s^\prime) \gamma^\mu u (p, s) 
\bar{u} (k^\prime, \lambda^\prime) \gamma^\nu u (k, \lambda).
\eeqar{M_BT}

\medskip

In this paper we consider the case of highly energetic jets, 
where $|\vec{\mathbf{q}}| \ll E$. In this limit $E^\prime$ becomes 
$E^\prime \approx E-\vec{\mathbf{v}}\cdot\vec{\mathbf{q}}$. Here 
$\vec{\mathbf{v}}$ is the velocity of the initial jet, i.e. the jet
4-momentum $p$ is equal to 
$p=(\frac{M}{\sqrt{1-v^2}}, \frac{M \vec{\mathbf{v}}}{\sqrt{1-v^2}})$.

Further, the matrix element given in Eq.~(\ref{eq:2}) has to be squared, 
averaged over initial spin $s$ of the jet and summed over all other spins.
\beqar
\frac{1}{2} \sum_{spins} |M_{el}|^2= |J(p^\prime)|^2 \; \frac{1}{E^2} \; 
\frac{\sin[(\omega- \vec{\mathbf{v}}\cdot\vec{\mathbf{q}})\frac{L}{2 v}]^2}
{(\omega- \vec{\mathbf{v}}\cdot \vec{\mathbf{q}})^2} \; 
\frac{1}{2} \sum_{spins} |{\cal M}|^2.
\eeqar{M_square}
where $\frac{1}{2} \sum_{spins} |{\cal M}|^2$ is given in Appendix B.

\bigskip

The differential energy loss is equal to $dE_{el}= \omega \,d\Gamma_{el}$, 
where collisional interaction rate $d\Gamma_{el}$ can be extracted from 
Eq.~(\ref{M_square}) as (see~\cite{GLV})
\beqar
d^{3}N_J \; d\Gamma_{el} \; &\approx& \frac{1}{2} \sum_{spins} |M_{el}|^{2} 
\frac{d^{3} 
\vec{{\bf p^\prime}}}{(2 \pi)^{3} 2 E^\prime }
\frac{d^{3} \vec{{\bf k}}}{(2 \pi)^{3} 2 k } 
\frac{d^{3} \vec{{\bf k^\prime}}}{(2 \pi)^{3} 2 k^\prime }
\sum_{\xi=q, \bar{q}, g} n_{eq}^\xi (k) 
(1\pm n_{eq}^\xi (k^\prime) ) . 
\eeqar{imm1a} 
Here
\beqar
d^{3} N_{J} = d_{R} |J(p^\prime)|^{2} 
\frac{d^{3}\vec{\mathbf{p^\prime}}}{( 2\pi )^{3} 2 E^\prime} \; ,
\eeqar{FO2}
and $d_R=3$ (for three dimensional representation of the quarks). In 
Eq.~(\ref{imm1a}), $n_{eq}^\xi (k)$ is the equilibrium momentum distribution 
at temperature $T$ of the incoming parton $\xi$, where $\xi$ denotes quark, 
antiquark or gluon. $(1\pm n_{eq}^\xi (k^\prime))$ is a factor associated 
with the outgoing parton, where $+$ corresponds to the gluon contribution, 
and $-$ to (anti)quark contribution. 

In this paper, we are interested only in the computation of the collisional 
energy loss. It is straightforward to show that in the collisional energy 
loss calculations, the $\pm n_{eq}^\xi (k^\prime)$ part in 
$(1\pm n_{eq}^\xi (k^\prime))$ can be dropped because the corresponding term 
in the energy loss integrand is odd under the exchange of $\vec{{\bf k}}$ 
and $\vec{{\bf k^\prime}}$, and integrates to zero (see 
also~\cite{BT_fermions}). Therefore, for the purpose of computing the 
collisional energy loss, we can replace $(1 \pm n_{eq}^\xi (k^\prime))$ by 
$1$ in Eq.~(\ref{imm1a}), which leads to
\beqar
d^{3}N_J \; d\Gamma_{el} \; &\approx& 
\frac{1}{2} \sum_{spins} |M_{el}|^{2} 
\frac{d^{3} 
\vec{{\bf p^\prime}}}{(2 \pi)^{3} 2 E^\prime }
\frac{d^{3} \vec{{\bf k}}}{(2 \pi)^{3} 2 k } 
\frac{d^{3} \vec{{\bf k^\prime}}}{(2 \pi)^{3} 2 k^\prime }
n_{eq} (k). 
\eeqar{imm1aELoss} 
Here, $n_{eq}(k)=\sum_{\xi=q, \bar{q}, g} n_{eq}^\xi (k)$  
is the equilibrium momentum distribution~\cite{BT_fermions} at temperature 
$T$ including quark, antiquark and gluon contributions (see Eq.~(\ref{n_eq})).
  
\medskip

The collisional energy loss can now be obtained from 
Eqs.~(\ref{M_square}), (\ref{FO2}) and (\ref{imm1aELoss}), leading to
\beqar
\Delta E_{el} \approx C_R \frac{1}{E^2} \int \frac{d^{3} 
\vec{{\bf k}}}{(2 \pi)^{3} 2 k} 
n_{eq} (k)  \int \frac{d^{3} \vec{{\bf k^\prime}}}{(2 \pi)^{3} 2 k^\prime }\;  
\omega \; 
\frac{\sin[(\omega- \vec{\mathbf{v}}\cdot\vec{\mathbf{q}})\frac{L}{2 v}]^2}
{(\omega- \vec{\mathbf{v}}\cdot \vec{\mathbf{q}})^2} \; 
\frac{1}{2} \sum_{spins} |{\cal M}|^2.
\eeqar{imm2a} 
Note that in Eq.~(\ref{imm2a}) we added a color factor $C_R$ and that 
$\vec{\mathbf{q}}=\vec{\mathbf{k}}^\prime-\vec{\mathbf{k}}$.

Equation~(\ref{imm2a}) can be further simplified by noting that, in a static 
medium, the collisional energy loss does not depend on the direction of 
$\vec{\mathbf{v}}$. After evaluating the $\frac{1}{2} \sum |{\cal M}|^2$ and 
averaging the integrand over the directions of $\vec{\mathbf{v}}$, we obtain 
(see Appendix B)
\beqar
\Delta E_{el} 
&=& \frac{C_R g^4}{2 \pi^4}  
\int_0^\infty n_{eq}(|\vec{\mathbf{k}}|) d |\vec{\mathbf{k}}| \; 
\left( \int_0^{|\vec{\mathbf{k}}|} |\vec{\mathbf{q}}|d |\vec{\mathbf{q}}| 
\int_{-|\vec{\mathbf{q}}|}^{|\vec{\mathbf{q}}|}\; \omega d \omega \;+
\int_{|\vec{\mathbf{k}}|}^{|\vec{\mathbf{q}}|_{max}} |\vec{\mathbf{q}}|d |\vec{\mathbf{q}}| 
\int_{|\vec{\mathbf{q}}|-2|\vec{\mathbf{k}}| }^{|\vec{\mathbf{q}}|}\; 
\omega d \omega \; \right)
 \nonumber \\
&& \hspace*{-2cm} \left( |\Delta_L(q)|^2 \frac{(2 |\vec{\mathbf{k}}|+\omega)^2  - 
|\vec{\mathbf{q}}|^2}{2} {\cal J}_1 +
|\Delta_T(q)|^2 \frac{(|\vec{\mathbf{q}}|^2-\omega^2)
((2 |\vec{\mathbf{k}}|+\omega)^2+ |\vec{\mathbf{q}}|^2)}
{4 |\vec{\mathbf{q}}|^4} \left[(v^2 |\vec{\mathbf{q}}|^2-\omega^2) {\cal J}_1 +
2 \omega {\cal J}_2- {\cal J}_3 \right] \right) \nonumber \\
\eeqar{Elastic_Eloss} 
where ${\cal J}_1$, ${\cal J}_2$ and ${\cal J}_3$ are given by 
Eqs.~(\ref{J1}-\ref{J3}) in Appendix B, and $|\vec{\mathbf{q}}|_{max}$ is given
by the following formula~\cite{TG}

\beqar
|\vec{\mathbf{q}}|_{max}=\text{Min}[E, \frac{2 |\vec{\mathbf{k}}| 
(1-|\vec{\mathbf{k}}|/E)}{1-v+2|\vec{\mathbf{k}}|/E}].
\eeqar{q_max}

Further numerical study of the collisional energy loss in a finite size QCD
medium is given in Section 4.

\section{Collisional energy loss in infinite QCD medium} 

Previous calculations of the collisional energy loss, in 
particular~\cite{Mustafa,TG,BT} were done for an infinite QCD medium. A problem 
with these calculations was that they produce unphysical energy gain in the 
lower momentum regions. Additionally, these computations leaded to different 
results and consequently provided quite a large uncertainty in the heavy quark 
(especially bottom) collisional energy loss.

In this Section and Appendix B.1 we present an improved calculation of 
collisional energy loss in an infinite QCD medium, with the goal of 1) 
removing the problems associated with previous calculations and 2) producing 
reliable infinite medium results which we will in Section 4 compare with the 
collisional energy loss in a finite medium.

In the case of an infinite QCD medium, the collisional energy loss per unit 
length $\frac{d E_{el}}{d L}$ is computed by assuming that the jet is produced 
at $x_0=-\infty$. The energy loss for a finite size medium is than 
(simplistically) calculated by multiplying this $\frac{d E_{el}}{d L}$ with 
the thickness $L$ of the medium.

In Appendix~B.1 we present an improved calculation of the collisional energy 
loss per unit length in an infinite QCD medium. The following result is 
obtained:

\beqar
\frac{d E_{el}}{d L} &=&\frac{g^4}{6 \, v^2 \,\pi^3}  
\int_0^\infty n_{eq}(|\vec{\mathbf{k}}|) d |\vec{\mathbf{k}}| \; 
\left( \int_0^{|\vec{\mathbf{k}}|/(1+v)} d |\vec{\mathbf{q}}| 
\int_{-v |\vec{\mathbf{q}}|}^{v |\vec{\mathbf{q}}|}\; \omega d \omega \;+
\int_{|\vec{\mathbf{k}}|/(1+v)}^{|\vec{\mathbf{q}}|_{max}} d |\vec{\mathbf{q}}|
\int_{|\vec{\mathbf{q}}|-2|\vec{\mathbf{k}}| }^{v |\vec{\mathbf{q}}|}\; 
\omega d \omega \; \right) \nonumber \\
&& \hspace*{-1cm}\left( |\Delta_L(q)|^2 \frac{(2 |\vec{\mathbf{k}}|+\omega)^2  
- |\vec{\mathbf{q}}|^2}{2}  + 
|\Delta_T(q)|^2 \frac{(|\vec{\mathbf{q}}|^2-\omega^2)
((2 |\vec{\mathbf{k}}|+\omega)^2+ |\vec{\mathbf{q}}|^2)}
{4 |\vec{\mathbf{q}}|^4} (v^2 |\vec{\mathbf{q}}|^2-\omega^2) \right).
\eeqar{Eel_infinite} 

To compare our result with the computations done in~\cite{TG,BT}, we here
introduce an arbitrary intermediate momentum scale 
$|\vec{\mathbf{q}}|^*$~\cite{BT}. The contribution from 
$|\vec{\mathbf{q}}|<|\vec{\mathbf{q}}|^*$ is denoted {\it soft}, while
contribution from $|\vec{\mathbf{q}}|>|\vec{\mathbf{q}}|^*$ is denoted
{\it hard}~\cite{BT}.

The soft contribution is given by
\beqar
\frac{d E_{el}^{soft}}{d L} &=&\frac{g^4}{6 \, v^2 \,\pi^3}  
\int_0^\infty n_{eq}(|\vec{\mathbf{k}}|) d |\vec{\mathbf{k}}| \; 
\int_0^{|\vec{\mathbf{q}}|^*} d |\vec{\mathbf{q}}| 
\int_{-v |\vec{\mathbf{q}}|}^{v |\vec{\mathbf{q}}|}\; \omega d \omega \;
\nonumber \\
&& \hspace*{-1cm}\left( |\Delta_L(q)|^2 \frac{(2 |\vec{\mathbf{k}}|+\omega)^2  - 
|\vec{\mathbf{q}}|^2}{2}  + |\Delta_T(q)|^2 \frac{(|\vec{\mathbf{q}}|^2-\omega^2)
((2 |\vec{\mathbf{k}}|+\omega)^2+ |\vec{\mathbf{q}}|^2)}
{4 |\vec{\mathbf{q}}|^4} (v^2 |\vec{\mathbf{q}}|^2-\omega^2) \right).
\eeqar{Eel_soft1} 
while the hard contribution is given by

\beqar
\frac{d E_{el}^{hard}}{d L} &=&\frac{g^4}{6 \, v^2 \,\pi^3}  
\int_0^\infty n_{eq}(|\vec{\mathbf{k}}|) d |\vec{\mathbf{k}}| \; 
\left( \int_{|\vec{\mathbf{q}}|^*}^{|\vec{\mathbf{k}}|/(1+v)} d|\vec{\mathbf{q}}| 
\int_{-v |\vec{\mathbf{q}}|}^{v |\vec{\mathbf{q}}|}\; \omega d \omega \;+
\int_{|\vec{\mathbf{k}}|}^{|\vec{\mathbf{q}}|_{max}} d |\vec{\mathbf{q}}| 
\int_{|\vec{\mathbf{q}}|-2|\vec{\mathbf{k}}| }^{v |\vec{\mathbf{q}}|}\; 
\omega d \omega \; \right)
\nonumber \\
&& \hspace*{-1cm}\left( |\Delta_L(q)|^2 \frac{(2 |\vec{\mathbf{k}}|+\omega)^2  - 
|\vec{\mathbf{q}}|^2}{2}  + |\Delta_T(q)|^2 \frac{(|\vec{\mathbf{q}}|^2-\omega^2)
((2 |\vec{\mathbf{k}}|+\omega)^2+ |\vec{\mathbf{q}}|^2)}
{4 |\vec{\mathbf{q}}|^4} (v^2 |\vec{\mathbf{q}}|^2-\omega^2) \right).
\eeqar{Eel_hard1} 

The soft contribution can be further simplified by keeping only the even 
contributions in the $\omega$ integral (the odd contributions vanish under 
symmetric integration)
 
\beqar
\frac{d E_{el}^{soft}}{d L} &=&\frac{g^4}{3 \, v^2 \,\pi^3}  
\int_0^\infty |\vec{\mathbf{k}}| n_{eq}(|\vec{\mathbf{k}}|) 
d |\vec{\mathbf{k}}| \; \int_0^{|\vec{\mathbf{q}}|^*} 
d |\vec{\mathbf{q}}| \nonumber \\
&& \hspace*{1.15cm}\int_{-v |\vec{\mathbf{q}}|}^{v |\vec{\mathbf{q}}|}\; 
\omega^2 d \omega \;\left( |\Delta_L(q)|^2 + 
\frac{1}{2} (1-\frac{\omega^2}{|\vec{\mathbf{q}}|^2})
(v^2 - \frac{\omega^2}{|\vec{\mathbf{q}}|^2}) |\Delta_T(q)|^2  \right)\nonumber 
\\ &=& \frac{g^2}{6 \,\pi\, v^2 } \; \mu_D^2  
\int_0^{|\vec{\mathbf{q}}|^*} 
d |\vec{\mathbf{q}}| \int_{-v |\vec{\mathbf{q}}|}^{v |\vec{\mathbf{q}}|}\; 
\omega^2 d \omega \;\left( |\Delta_L(q)|^2 + 
\frac{1}{2} (1-\frac{\omega^2}{|\vec{\mathbf{q}}|^2})
(v^2 - \frac{\omega^2}{|\vec{\mathbf{q}}|^2}) |\Delta_T(q)|^2  \right)
\eeqar{Eel_soft} 
where in the last step we used the fact that
\beqar
\int_0^\infty |\vec{\mathbf{k}}| n_{eq}(|\vec{\mathbf{k}}|) d |\vec{\mathbf{k}}|
=\frac{\pi^2 T^2}{2}(1+\frac{N_f}{6}).
\eeqar{Int_neq}

Equation~(\ref{Eel_soft}) agrees with the Eq.~(55) in~\cite{BT_fermions}.

\bigskip 

For the hard contribution we use that, in the limit of large momentum transfer 
($|\vec{\mathbf{q}}| \gg |\vec{\mathbf{q}}|^*$), 
$|\Delta_L(q)|\rightarrow \frac{1}{|\vec{\mathbf{q}}|^2}$ and
$|\Delta_T(q)|\rightarrow \frac{1}{\omega^2-|\vec{\mathbf{q}}|^2}$. It is than 
straightforward to show that the hard contribution reduces to the Eq.~(17) 
in~\cite{BT_fermions}, i.e. 
\beqar
\frac{d E_{el}^{hard}}{d L} &=&\frac{g^4}{6 \, v^2 \,\pi^3}  
\int_0^\infty n_{eq}(|\vec{\mathbf{k}}|) d |\vec{\mathbf{k}}| \; 
\left( \int_{|\vec{\mathbf{q}}|^*}^{2|\vec{\mathbf{k}}|/(1+v)} 
\frac{d |\vec{\mathbf{q}}|}{|\vec{\mathbf{q}}|^2}  
\int_{-v |\vec{\mathbf{q}}|}^{v |\vec{\mathbf{q}}|}\; \omega d \omega \;+
\int_{2|\vec{\mathbf{k}}|/(1+v)}^{|\vec{\mathbf{q}}|_{max}} 
\frac{d |\vec{\mathbf{q}}|}{|\vec{\mathbf{q}}|^2} 
\int_{|\vec{\mathbf{q}}|-2|\vec{\mathbf{k}}| }^{v |\vec{\mathbf{q}}|}\; 
\omega d \omega \; \right)
\nonumber \\
&&\left( \frac{3\, \omega^2}{4 |\vec{\mathbf{q}}|^2}- \frac{v^2}{4} - 
\frac{1-v^2}{2}\frac{|\vec{\mathbf{q}}|^2}{|\vec{\mathbf{q}}|^2-\omega^2}+
3\frac{|\vec{\mathbf{k}}|(|\vec{\mathbf{k}}|+\omega)}{|\vec{\mathbf{q}}|^2}-
(1-v^2)\frac{|\vec{\mathbf{k}}|(|\vec{\mathbf{k}}|+\omega)}
{|\vec{\mathbf{q}}|^2-\omega^2} \right).
\eeqar{Eel_hard} 
Equations~(17) and~(55) from~\cite{BT_fermions} were used in Ref.~\cite{BT} 
to obtain their Eqs.~(8) and~(12). That is, while our Eq.~(\ref{Eel_infinite}) 
is more general, in special cases (i.e. Eqs.~(\ref{Eel_soft}) 
and~(\ref{Eel_hard})) it reproduces results from~\cite{BT}. 

\medskip 

The computation in~\cite{TG} considered only the soft gluon limit, and 
replaced $|\vec{\mathbf{q}}|^*$ by $|\vec{\mathbf{q}}|_{max}$, where 
$|\vec{\mathbf{q}}|_{max}$ is given by Eq.~(\ref{q_max}). Consequently, for 
the purpose of comparison with~\cite{TG}, we replaced $|\vec{\mathbf{q}}|^*$ 
by $|\vec{\mathbf{q}}|_{max}$ in Eq.~(\ref{Eel_soft}). Additionally, the 
problem with this approach is that, in the high momentum $|\vec{\mathbf{q}}|$ 
region, the method~\cite{TG} is not able to treat the lower $\omega$ bound 
properly (compare Eq.~(\ref{Eel_soft}) with Eq.~(\ref{Eel_infinite}) where 
$\omega$ bounds are properly treated). To overcome this problem, the 
calculation in~\cite{TG} was limited to the forward emission only (i.e. 
$\omega>0$). If this limit is also taken into account, our Eq.~(\ref{Eel_soft}) 
reproduces Eq.(4.1) in~\cite{TG}.
 
\medskip 

In summary, in the known limiting cases, our result (i.e. 
Eq.~(\ref{Eel_infinite})) reduces to the results published in~\cite{TG,BT}. 
The advantage of our result over~\cite{TG} is that it includes the hard 
contribution and consistently treats the integration limits. Comparing 
to~\cite{BT}, the advantage of our result is that it does not make a sharp 
transition from soft to hard limits, and consequently it does not require the 
introduction of an unphysical momentum scale $|\vec{\mathbf{q}}|^*$ as 
in~\cite{BT}. 

\section{Numerical results}

In this section we give a numerical study of the collisional energy loss in 
a QCD medium as presented in Sections~3 and~4. To do this, we further assume 
that the QCD plasma is characterized by $T=0.225$~GeV, $N_f=2.5$ and 
$\alpha=0.3$. For the light quark mass we take $M=\mu_D/\sqrt{6}$, where 
$\mu_D=g^2 T^2 (1+\frac{N_f}{6}) \approx 0.5$~GeV is the Debye mass. The 
charm mass is taken to be $M=1.2$~GeV, and the bottom mass is $M=4.75$~GeV. 

\subsection{Collisional energy loss in infinite QCD medium}

\begin{figure}[h]
\vspace*{5.2cm} \includegraphics{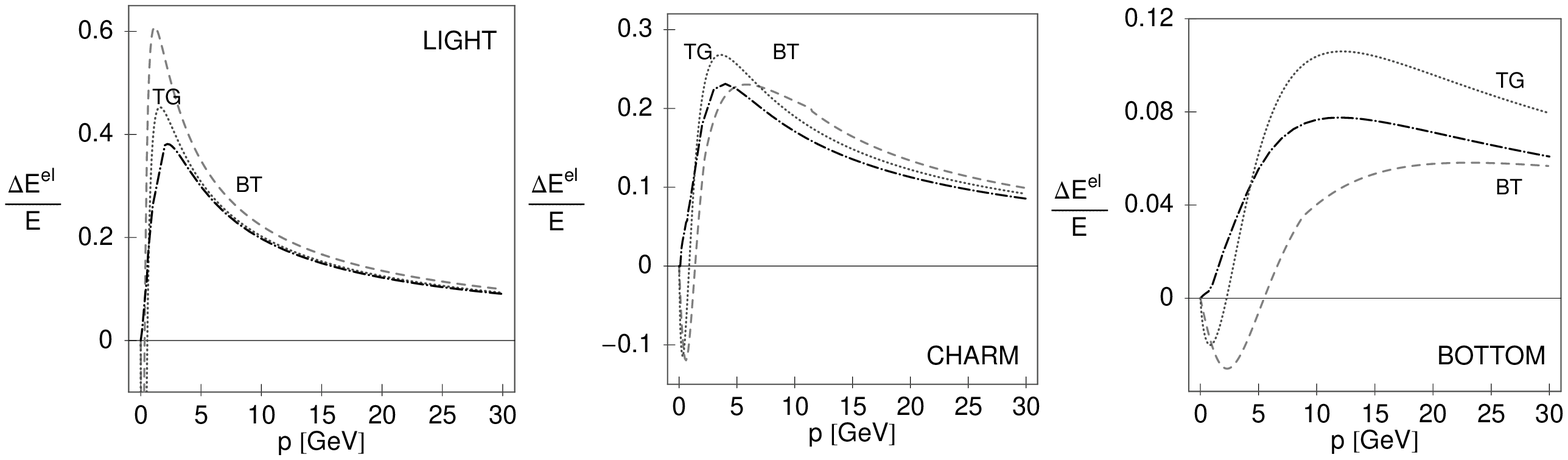}
\caption{ Fractional collisional energy loss is shown as a function of 
momentum for light, charm and bottom quark jets (left, center and right panels 
respectively). Dash-dotted curves are obtained by using 
Eq.~(\ref{Eel_infinite}) from this paper. Dashed curves correspond to 
Eqs.~(8) and~(12) from~\cite{BT}, while dotted curves are obtained by using 
Eq.~(4.1) from~\cite{TG}. Assumed thickness of the medium is $L=5$~fm.}
\label{D_TG_BT}
\end{figure}

In Fig.~\ref{D_TG_BT} we compare our collisional energy loss results in an
infinite QCD medium (Eq.~(\ref{Eel_infinite})) with previous computations 
by~\cite{TG,BT}. We see that, while both BT~\cite{BT} and TG~\cite{TG} 
computations lead to unphysical negative energy loss results in the low 
momentum region, our computations give positive collisional energy loss in the 
whole jet momentum range. This is particularly important in the bottom quark 
case where the unphysical behavior persists up to 5~GeV in BT~\cite{BT} case 
and up to 2~GeV in TG~\cite{TG} case. The reason for this behavior is that only
the leading logarithmic contribution was considered in the final steps of both 
BT and TG computations. Note that the problem of unphysical energy gain was 
addressed in Ref.~\cite{Romatschke}, by including fully dressed gluon propagator 
in their calculations. However, that approach leaded to another problem, since 
the unphysical momentum scale $|\vec{\mathbf{q}}|^*$ appeared in the collisional 
energy loss results~\cite{Romatschke}. Therefore, our results present a first 
complete solution to the unphysical energy gain problem.

\medskip

Our numerical results agree with BT only in the asymptotic regions, which is 
likely the consequence of the following: 1) BT made a sharp (instead of 
continuous) transition from soft to hard limit and 2) they introduced a sharp 
boundary in the energy loss computations depending on whether the initial jet 
energy is much larger/smaller than $M^2/T$. 

\medskip

Despite the fact that the BT computations are more up to date and treat the 
collisional energy loss more consistently than TG, we see that our results 
show better agreement with TG~\cite{TG} computations. Particularly, in case of
light and charm quark jets, there is a quite good agreement between our
results and those of TG~\cite{TG}. The good agreement is probably because the 
forward emission only (see Section~3) provides a quite plausible estimate for 
the collisional energy loss. However, for bottom quarks we see that neither BT 
nor TG computations provide a reasonable estimate for the collisional energy 
loss. Therefore, in this case, the more accurate computation of collisional 
energy loss (i.e. our Eq.~(\ref{Eel_infinite})) is needed.

\subsection{Collisional energy loss in a finite QCD medium}

\begin{figure}[h]
\vspace*{5.2cm} \includegraphics{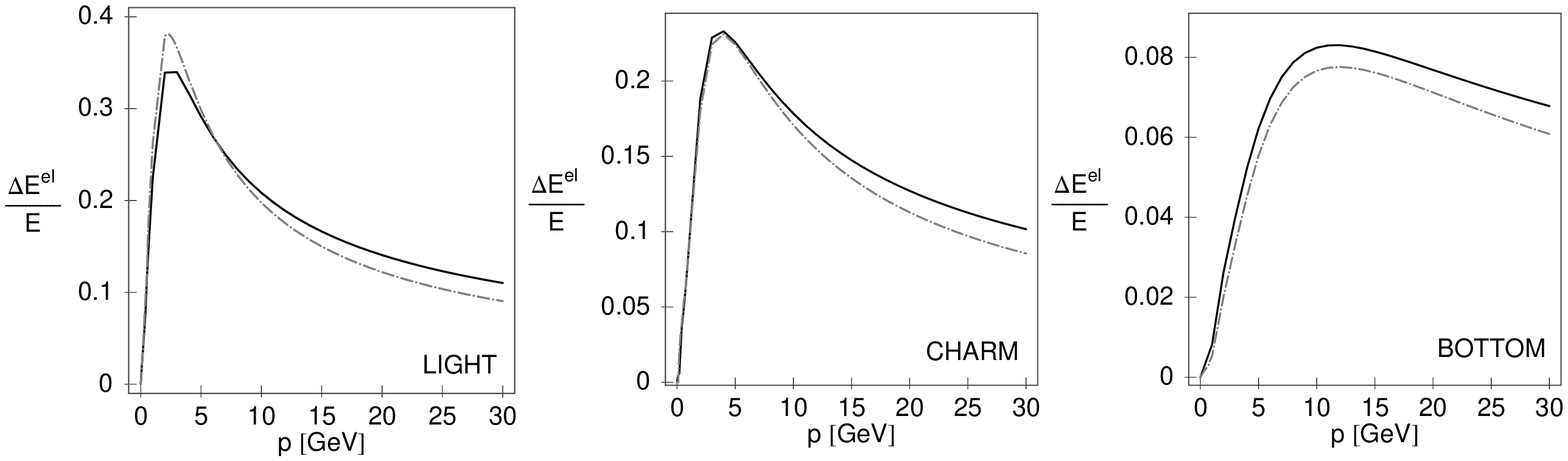}
\caption{ Fractional collisional energy loss is shown as a function of 
momentum for light, charm and bottom quark jets (left, center and right panels 
respectively). Full curves correspond to finite medium case (see 
Eq.~(\ref{Elastic_Eloss})), while dash-dotted curves correspond to 
infinite medium case (see Eq.~(\ref{Eel_infinite})). Assumed thickness of the 
medium is $L=5$~fm.}
\label{E_compare}
\end{figure}

\begin{figure}[h]
\vspace*{7cm} \includegraphics{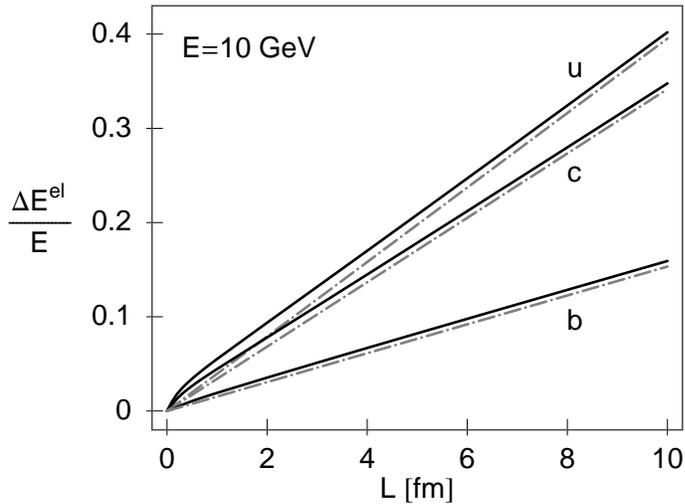}
\caption{ Fractional collisional energy loss is shown as a function of 
thickness of the medium for light, charm and bottom quark jets (upper, middle 
and lower set of curves respectively). Full curves correspond to finite medium 
case (see Eq.~(\ref{Elastic_Eloss})), while dash-dotted curves correspond to 
infinite medium case (see Eq.~(\ref{Eel_infinite})). Initial momentum of the 
jet is 10 GeV.}
\label{L_compare}
\end{figure}

\begin{figure}
\vspace*{7.cm} \includegraphics{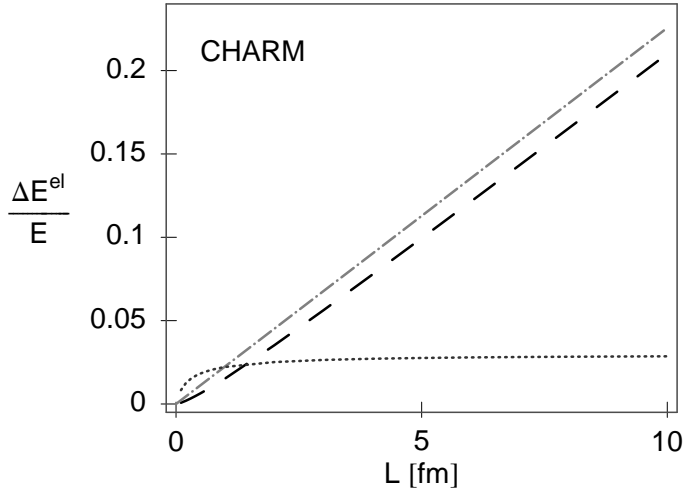}
\caption{ Fractional collisional energy loss is shown as a function of 
thickness of the medium for charm quark jet. Dash-dotted curves correspond to 
infinite medium case (see Eq.~(\ref{Eel_infinite})), while dashed curve show 
what would be the collisional energy loss in finite size medium if term  
$2 \omega {\cal J}_2-{\cal J}_3 = 0$. Dotted curve shows the extra contribution
to the collisional energy loss in finite size medium, due the fact that 
$2 \omega {\cal J}_2-{\cal J}_3 \ne 0$. Initial momentum of the jet is 20 GeV.}
\label{Extra_L}
\end{figure}

Figures~\ref{E_compare} and~\ref{L_compare} show the comparison between 
collisional energy loss in infinite and finite size QCD medium. Contrary 
to~\cite{Peigne}, we find that a finite size medium does not have a large 
effect on the collisional energy loss. The discrepancy between our results and
those presented in~\cite{Peigne} is due to the fact that what was called 
collisional energy loss in~\cite{Peigne}, is in fact combination of collisional
and part of the $0^{th}$ order radiative energy loss. Actually, the calculation 
in~\cite{Peigne} does not present a complete $0^{th}$ order energy loss either, 
since transition radiation~\cite{MD_TR} was not included in their computation.

\medskip

Contrary to naive expectations, from Figs.~\ref{E_compare} and~\ref{L_compare} 
we found that collisional energy loss in a finite size medium can be somewhat 
larger than in an infinite medium. The reason is that in 
Eq.~(\ref{Eel_infinite}) there exists a term $2 \omega {\cal J}_2-{\cal J}_3$. 
If this term were equal to zero (as in the case of infinite medium), the 
energy loss in a finite medium case would always be smaller than in an 
infinite medium, as naively expected (compare dashed and dot-dashed curves in 
Fig.~\ref{Extra_L}). However, in the finite medium case, the term 
$2 \omega {\cal J}_2-{\cal J}_3 \ne 0$, giving a noticeable positive 
contribution (see the dotted curve in Fig.~\ref{Extra_L}) which will lead to 
somewhat larger (overall) energy loss in the finite medium case.

\medskip

To further discuss this, let us look at the Eqs.~(\ref{J1}-\ref{J3}) 
(Appendix A) in the finite medium case, and compare them to the corresponding 
Eqs.~(\ref{J1_limit}-\ref{J3_limit}) in the infinite medium. The $\delta$ 
function in Eqs.~(\ref{J1_limit}-\ref{J3_limit}) ensures energy conservation, 
which is satisfied when the jet is produced at $-\infty$. Consequently, in 
this case $2 \omega {\cal J}_2-{\cal J}_3 \equiv 0$ (see Eqs.~(\ref{J2_limit}) 
and~\ref{J3_limit})). However, when the jet is produced at finite time $x_0$, 
time translation invariance is broken, and therefore the energy of the 
collisional process is not conserved, leading to 
$2 \omega {\cal J}_2-{\cal J}_3 \ne 0$.


\subsection{Comparison between collisional and radiative energy loss in a
finite size QCD medium}

In Appendix A we showed how to separate the contributions to the collisional 
and radiative energy loss. In this section we directly compare these two 
contributions in the case of finite size QCD medium. 

\begin{figure}[h]
\vspace*{5.3cm} \includegraphics{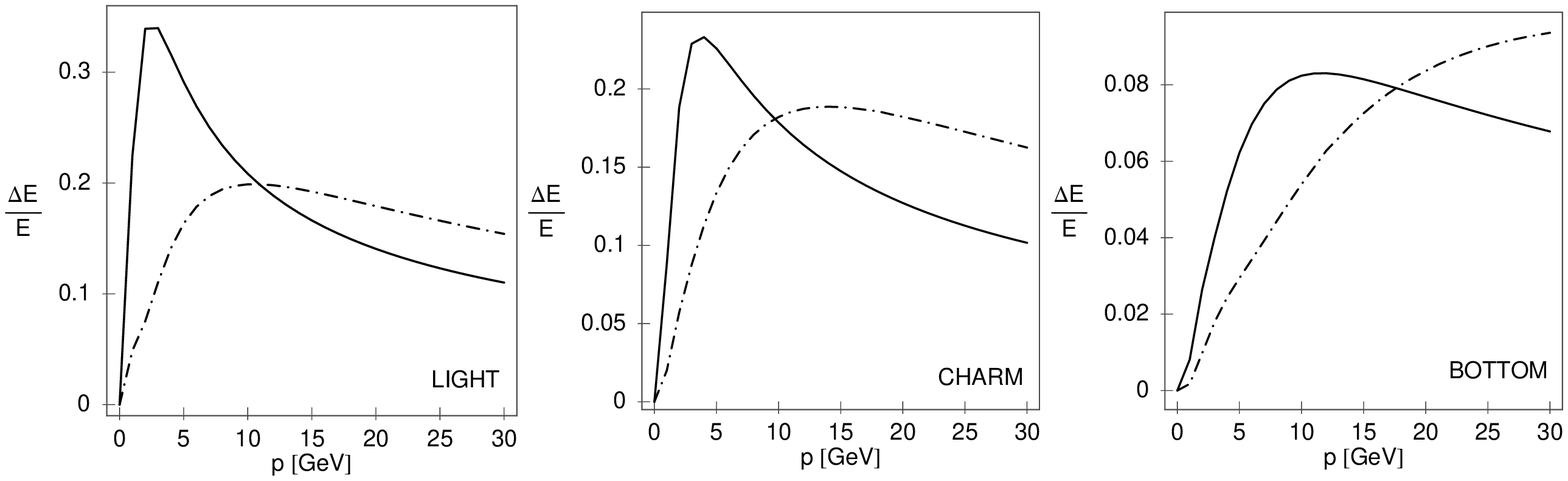}
\caption{ The comparison between collisional and radiative fractional energy 
loss is shown as a function of momentum for light, charm and bottom quark jets 
(left, center and right panels respectively). Full curves show the collisional 
energy loss, while dot-dashed curves show the net radiative energy loss. 
Assumed thickness of the medium is $L=5$~fm and $\lambda=1.2$~fm~\cite{WHDG}.}
\label{Comparison_E}
\end{figure}

\begin{figure}
\vspace*{5.3cm} \includegraphics{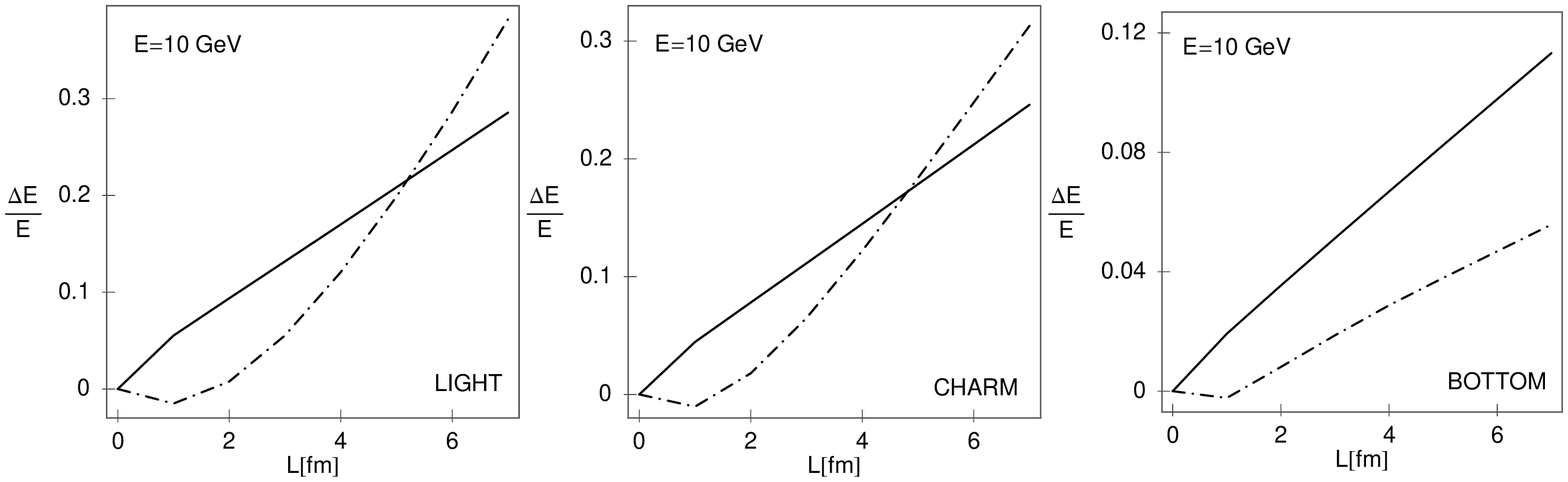}
\caption{ The comparison between collisional and radiative fractional energy 
loss is shown as a function of the thickness of the medium. Light, charm and 
bottom quark cases are shown on the left, center and right panels respectively.
Full curves show the collisional energy loss, while dot-dashed curves show the 
net radiative energy loss. Mean free path is $\lambda=1.2$~fm~\cite{WHDG}. 
Initial momentum of the jet is $E=10$ GeV. }
\label{Comparison_L}
\end{figure}

To compute the net radiative energy loss, we note that there are three 
important effects that control this energy loss in a QCD medium. These effects
are the Ter-Mikayelian effect~\cite{DG_TM}, transition radiation~\cite{MD_TR} 
and medium induced radiation~\cite{DG_Ind}. In~\cite{MD_TR}, we combined these 
effects to compute the net radiative energy loss. We here use these results for
the purpose of further comparison with the collisional energy loss. Note 
that in these computations, in order to simulate confinement in the vacuum, 
we introduce an effective gluon mass in the vacuum 
$m_{g,v}\approx \Lambda_{QCD}=0.2$ GeV (for more details see~\cite{DG_TM}).

\medskip

In Figures~\ref{Comparison_E} and~\ref{Comparison_L} we show the collisional 
and radiative energy loss as a function of jet energy and thickness of the 
medium, respectively. We see that collisional energy loss is comparable with 
the net radiative energy loss in the medium. Therefore, the collisional energy 
loss contribution is significant and must be included in the computation of 
jet quenching in a QCD medium.

In particular, we note that in the lower momentum (i.e. $p<10$ GeV) range, the
collisional energy loss dominates the radiative one. At RHIC, jet suppression 
is mostly measured in this (lower) momentum range. Therefore, contrary to 
previous studies~\cite{MVWZ:2004}-\cite{KW:2004}, our results indicate that it 
is collisional instead of radiative energy loss which gives the dominant 
contribution to the observed suppression values.

\section{Conclusion}

This paper addressed the $0^{th}$ order contribution to the collisional energy 
loss in a finite size QCD medium. The interest in the collisional energy loss 
has been renewed by the recent studies~\cite{Mustafa,Dutt-Mazumder}, 
particularly in the context of the single electron 
puzzle~\cite{Adler:2005xv}-\cite{Djordjevic:2005db}. In 
Refs.~\cite{Mustafa,Dutt-Mazumder} it was claimed that, contrary to the 
previous beliefs, for the parameter ranges relevant in URHIC, radiative and 
collisional energy loss become comparable. However, a recent
study by Peigne {\em et al.}~\cite{Peigne} suggested that collisional energy 
loss is large in the ideal infinite medium case, while the finite size medium 
effects lead to significant reduction of the collisional energy loss. The 
paper~\cite{Peigne}, however, did not completely separate collisional from 
radiative energy loss. 

\medskip

Additionally, even in the infinite medium case, the problem of collisional 
energy loss was not completely solved. Previous computations obtained 
unphysical results in the low momentum regions~\cite{TG,BT}, and an approach 
to solve this problem~\cite{Romatschke} leaded to the results dependent on the 
unphysical momentum scales. In addition, these computations introduced quite 
a large uncertainty in the heavy quark (especially bottom) collisional energy 
loss, since they leaded to noticeably different numerical results.

\medskip

The above reasons and the previously discussed single electron puzzle, 
motivated us to provide a detailed study of the $0^{th}$ order collisional 
energy loss in a finite size QCD medium created in URHIC. First, in Appendix A 
we showed that, though $0^{th}$ order collisional and radiative energy loss 
contributions come from the same one-loop HTL diagram, there is no overlap 
between collisional and radiative energy loss computations. More specifically, 
while $0^{th}$ order collisional energy loss comes from the processes which 
have the same number of incoming and outgoing particles, the radiative energy 
loss has one gluon more as an outcome of the process. Additionally, we showed 
that in the $0^{th}$ order calculations, there are no interference effects 
between collisional and radiative energy loss, which is different from a result
in the recent paper~\cite{XNWang_coll}. The absence of interference effects 
comes from the fact that, contrary to~\cite{XNWang_coll}, in our study we 
consistently treat the gluon dispersion relation in the medium. This leads to 
the following conditions: 1) for the $0^{th}$ order collisional energy loss 
contributions, the energy of the exchanged (virtual) gluon has to be smaller, 
or equal, to the gluon momentum, and 2) for the radiative energy loss 
contributions the energy of the radiated gluon has to be larger than its 
momentum. Therefore, these two contributions take non-zero values in 
non-overlapping regions, and consequently cannot interfere with each other.  

\medskip

In the case of infinite medium, our computation interpolates smoothly between 
soft to hard contributions and, contrary to~\cite{BT}, does not require the 
introduction of an arbitrary intermediate momentum scale. Additionally, our 
computation treats the lower momentum region consistently, removing the 
unphysical energy gain results obtained in previous computations~\cite{TG,BT}.

\medskip

In the case of finite size QCD medium, contrary to the study by Peigne 
{\em et al.}~\cite{Peigne} we showed that finite size effects have small effect
on the collisional energy loss. Therefore, consistently with the claims in 
Refs.~\cite{Mustafa,Dutt-Mazumder} and our recent single electron suppression 
estimates~\cite{WHDG}, we here showed that collisional energy loss is 
important, and has to be included in the computation of jet quenching. 

\bigskip

\section*{Acknowledgments} I thank Ulrich Heinz for discussions and critical 
reading of the manuscript. Valuable discussions with Eric Braaten, Miklos 
Gyulassy, Yuri Kovchegov and Xin-Nian Wang are gratefully acknowledged. This 
work is supported by the U.S. Department of Energy, grant DE-FG02-01ER41190.

\begin{appendix}

\section{HTL contribution to the collisional energy loss}

\begin{figure}[h]
\vspace*{3.cm} \includegraphics{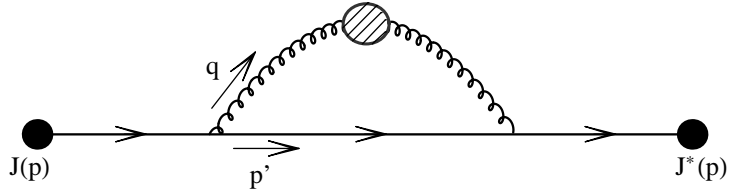}
\caption{One Hard Thermal Loop (HTL) diagram.}
\label{HTL_loop}
\end{figure}

In this section we will derive the formula for the lowest order collisional 
interaction rate. The zeroth order contribution to both radiative and 
collisional rates comes from the diagram $M$ given in Fig.~\ref{HTL_loop}. We 
will below use this diagram as a starting point to separate contributions of 
collisional and radiative energy loss.

\begin{figure}[h]
\vspace*{7.5cm}
\includegraphics{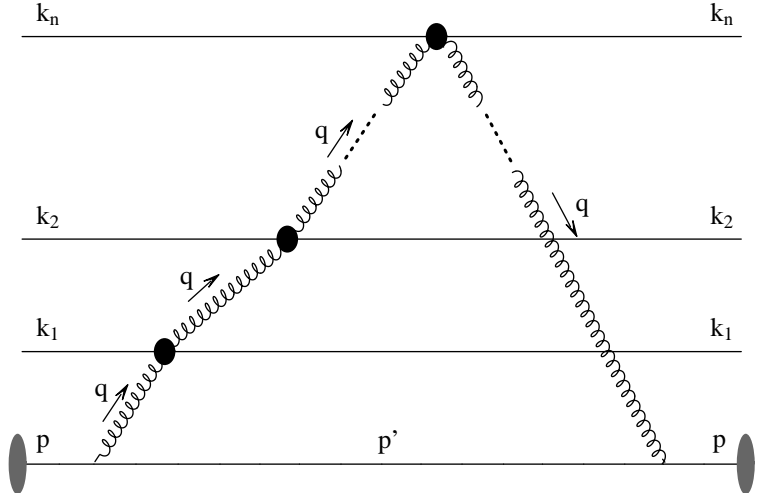}  
\caption{Diagram $M_n$}
\label{HTL_diag_Mn}
\end{figure}

Diagram $M$ corresponds to $\sum M_{n}$, where $M_{n}$ is the amplitude of 
the diagram shown in Fig.~\ref{HTL_diag_Mn}.

\begin{figure}[h]
\vspace*{3cm}
\includegraphics{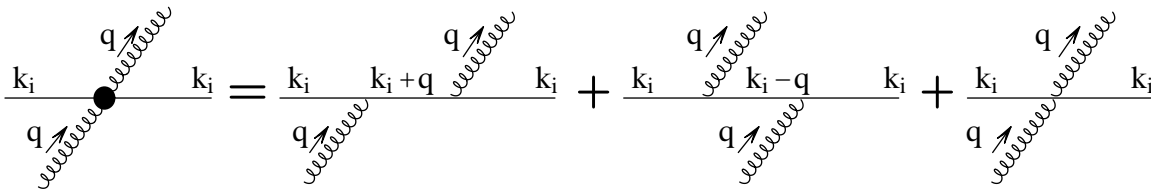}  
\vskip 0pt
\caption{Definition of the ``black circles'' in Diagram $M_n$.} 
\label{Black_circle}
\end{figure}

The definition of ``black circles'' in Fig.~\ref{HTL_diag_Mn} is shown in 
Fig.~\ref{Black_circle}.

\begin{figure}[h]
\vspace*{2.6cm} \includegraphics{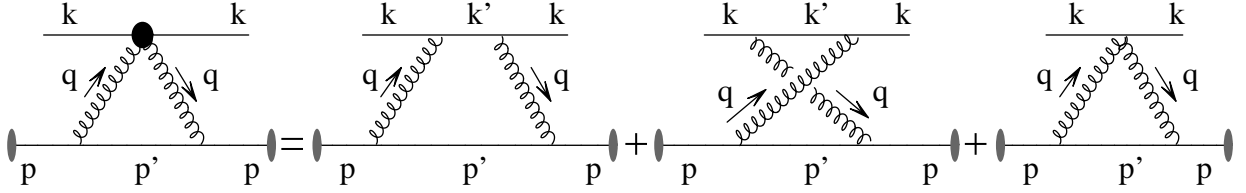}
\caption{ $n=1$ contribution to the HTL diagram.}
\label{Diag_n1}
\end{figure}

The diagram $M$ contains both collisional and radiative $0^{th}$ order 
contribution to the jet energy loss. It is useful to look at the simple $n=1$ 
case (see Fig.~\ref{Diag_n1}) to better understand this.

The contribution to the collisional energy loss is obtained by ``cutting'' 
(i.e. putting on shell) the propagators of parton $k'$ and $p'$. On the other
hand, the radiative contribution is obtained by putting the parton propagator 
$p'$ and the gluon propagator $q$ on shell. From this, it follows that 
collisional and radiative contributions come from different diagrams.
Furthermore, from the conservation of energy and momentum it can be shown that 
cutting the gluon propagator $q$, leads to the condition 
$|\omega|>|\vec{\mathbf{q}}|$, while cutting the propagator of parton 
$k^\prime$ leads to the condition $|\omega|<|\vec{\mathbf{q}}|$. Consequently, 
there is no interference (and over-counting) between collisional and radiative 
contributions\footnote{Note that in this paper, we treat only the  $0^{th}$ 
order contribution to collisional energy loss. It is, however, possible that
interference effects and/or over-counting between collisional and radiative 
energy loss contributions would occur in the higher order computations. 
Higher order contributions are a separate non-trivial problem, which is not 
considered in this paper.}. The computation of the 
radiative $0^{th}$ order energy loss has already been a subject of our previous 
work~\cite{MD_TR,DG_TM}. So, the contributions from the diagrams which give 
raise to the radiative energy loss, will not be further addressed here.

\begin{figure}[h]
\vspace*{2.5cm} 
\includegraphics{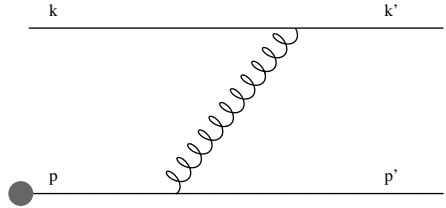}
\caption{Feynman diagram for the lowest order collisional energy loss.}
\label{ME0}
\end{figure}

As we can see from the right side of the Fig.~\ref{Diag_n1}, there are two 
contributions from diagram $M_1$ to the collisional rate. These two 
contributions can be combined into one by allowing that the energy of the 
gluon can take both positive and negative values. Therefore, the contribution 
to the collisional rate from diagram $M_1$ ($d\Gamma_{M_1}$) is equal to
\beqar
d^3 N_J \; d\Gamma_{M_1} =  \frac{d^{3} 
\vec{{\bf p^\prime}}}{(2 \pi)^{3} 2 E^\prime }
\frac{d^{3} \vec{{\bf k}}}{(2 \pi)^{3} 2 k } 
\frac{d^{3} \vec{{\bf k^\prime}}}{(2 \pi)^{3} 2 k^\prime }\; n_{eq} (k) \; 
|M_{E_0}|^2 \; ,
\eeqar{MEO_contribution}
where $M_{E_0}$ is the Feynman diagram shown in Fig.~\ref{ME0}, and 
$\omega \in (-|\vec{\mathbf{q}}|,|\vec{\mathbf{q}}|)$. Note that in the above 
equation each of the terms in $n_{eq} (k)=\sum_{\xi=q,\bar{q},g}n_{eq}^\xi(k)$ 
should be multiplied by an extra factor $(1 \pm n_{eq}^\xi (k^\prime))$ for 
the outgoing medium parton (see Section 2). Here $\xi$ 
corresponds to anti(quark) or gluon. The $+$ sign is associated with gluon and 
the $-$ sign with anti(quark) contributions. However, the 
$\pm n_{eq}^\xi (k^\prime)$ term in $(1 \pm n_{eq}^\xi (k^\prime))$ does not 
contribute to the collisional energy loss (for more details see Section 2 and 
Ref.~\cite{BT_fermions}). Therefore, in  
Eq.~(\ref{MEO_contribution}) we keep only the contributions that give raise 
to the collisional energy loss.

\begin{figure}[h]
\vspace*{6.5cm} \includegraphics{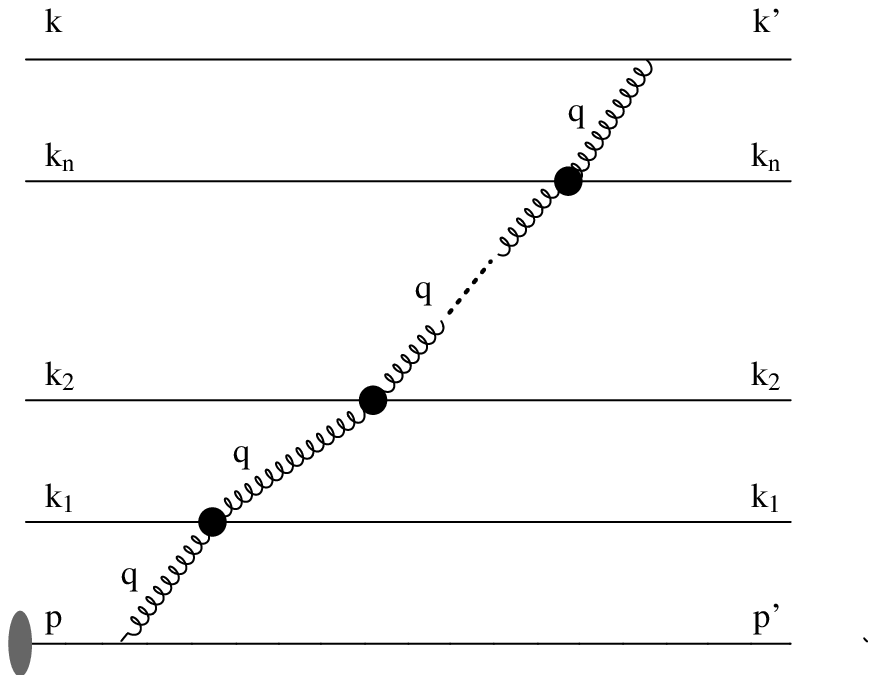}
\caption{Feynman diagram for the collisional energy loss with $n$ interactions 
with the medium.}
\label{MEn}
\end{figure}

In the same way, it can be shown that the contribution to the collisional 
energy loss from diagram $M_n$ is equal to 
\beqar
d^3 N_J \; d\Gamma_{M_n} = \frac{d^{3} 
\vec{{\bf p^\prime}}}{(2 \pi)^{3} 2 E^\prime }
\frac{d^{3} \vec{{\bf k}}}{(2 \pi)^{3} 2 k }  
\frac{d^{3} \vec{{\bf k^\prime}}}{(2 \pi)^{3} 2 k^\prime }\;n_{eq} (k)\;
\sum_{i=0}^{n-1} M_{E_i} M_{E_{n-1-i}}^\dagger \;,
\eeqar
where the diagram $M_{E_n}$ is shown in Fig.~\ref{MEn}.

Since $M=\sum_{n=0}^{\infty} M_n$, the contribution to the collisional energy 
loss from the diagram $M$ is equal to
\beqar
d^3 N_J \; d\Gamma &=&\frac{d^{3} 
\vec{{\bf p^\prime}}}{(2 \pi)^{3} 2 E^\prime }
\frac{d^{3} \vec{{\bf k}}}{(2 \pi)^{3} 2 k }
\frac{d^{3} \vec{{\bf k^\prime}}}{(2 \pi)^{3} 2 k^\prime }\; n_{eq} (k) \; 
\sum_{n=0}^{\infty} \sum_{i=0}^{n-1} M_{E_i} M_{E_{n-1-i}}^\dagger \nonumber \\
&=&\frac{d^{3} 
\vec{{\bf p^\prime}}}{(2 \pi)^{3} 2 E^\prime }
\frac{d^{3} \vec{{\bf k}}}{(2 \pi)^{3} 2 k }
\frac{d^{3} \vec{{\bf k^\prime}}}{(2 \pi)^{3} 2 k^\prime }\; n_{eq} (k) \;\sum_{n=0}^{\infty} \sum_{i=0}^{n} M_{E_i} M_{E_{n-i}}^\dagger \; .
\eeqar{ColM}

\begin{figure}[h]
\vspace*{2.5cm} \includegraphics{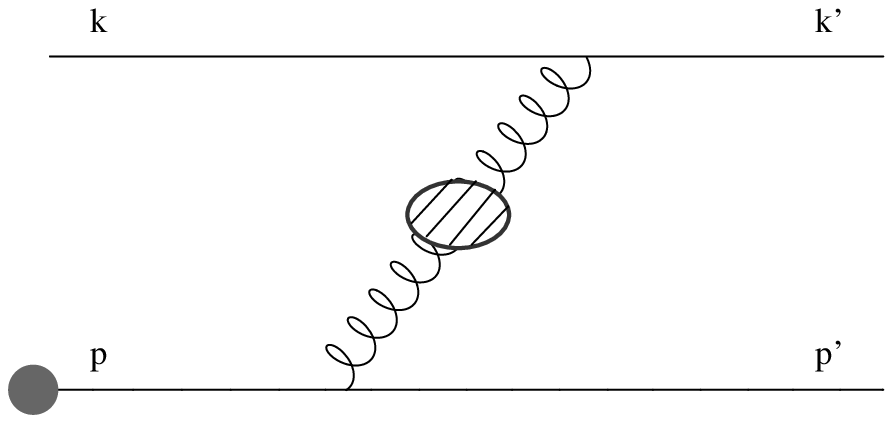}
\caption{Feynman diagram $M_{el}$ for the collisional energy loss in QCD 
medium. The large dashed circle (``blob'') represents the effective gluon 
propagator~\cite{DG_TM}.}
\label{ME}
\end{figure}

We next want to prove that $d^3 N_J \; d\Gamma =\frac{d^{3} 
\vec{{\bf p^\prime}}}{(2 \pi)^{3} 2 E^\prime }
\frac{d^{3} \vec{{\bf k}}}{(2 \pi)^{3} 2 k } 
\frac{d^{3} \vec{{\bf k^\prime}}}{(2 \pi)^{3} 2 k^\prime } \; n_{eq} (k) 
\; |M_{el}|^2$, where (see Fig.~\ref{ME})
\beqar
M_{el}=\sum_{n=0}^{\infty} M_{E_{n}}
\eeqar{M_E}

To prove the above, we will first compute $|M_{el}|^2$
\beqar
|M_{el}|^2=\sum_{i=0}^{\infty} \sum_{j=0}^{\infty} M_{E_i} M_{E_{j}}^\dagger=
\sum_{i=0}^{\infty} \sum_{n=i}^{\infty} M_{E_i} M_{E_{n-i}}^\dagger
\eeqar{M_Esq}
where in the last step we defined $n=i+j \rightarrow j=n-i$.

Since
\beqar
\sum_{i=0}^{\infty} \sum_{n=i}^{\infty} = \sum_{n=0}^{\infty} \sum_{i=0}^{n},
\eeqar{sum_eq}
we can conclude that 
\beqar
d^3 N_J \; d\Gamma =\frac{d^{3} 
\vec{{\bf p^\prime}}}{(2 \pi)^{3} 2 E^\prime }
\frac{d^{3} \vec{{\bf k}}}{(2 \pi)^{3} 2 k } 
\frac{d^{3} \vec{{\bf k^\prime}}}{(2 \pi)^{3} 2 k^\prime }\; n_{eq} (k) \;
\sum_{n=0}^{\infty} |M_{el}|^2 \; ,
\eeqar{Elastic_contribution}
which is what we wanted to prove. Therefore, the collisional interaction rate 
can be obtained by an intuitive approach of computing $|M_{el}|^2$ (see 
Fig.~\ref{ME}), where blob represents the effective gluon propagator.

\section{Collisional energy loss computations}

In this Appendix we will derive the collisional energy formula given by 
Eq.~(\ref{Elastic_Eloss}). To do that we start from the Eq.~(\ref{imm2a}), i.e.
\beqar
\Delta E_{el} = C_R \frac{1}{E^2} \int \frac{d^{3} 
\vec{{\bf k}}}{(2 \pi)^{3} 2 k} 
n_{eq} (k)  \int \frac{d^{3} \vec{{\bf k^\prime}}}{(2 \pi)^{3} 2 k^\prime } \;  
\omega \; 
\frac{\sin[(\omega- \vec{\mathbf{v}}\cdot\vec{\mathbf{q}})\frac{L}{2 v}]^2}
{(\omega- \vec{\mathbf{v}}\cdot \vec{\mathbf{q}})^2} \; 
\frac{1}{2} \sum_{spins} |{\cal M}|^2 
\eeqar{imm2a_App} 
where $n_{eq} (k)$ is the equilibrium momentum distribution~\cite{BT_fermions} at 
temperature $T$ including quarks and gluons
\beqar
n_{eq} (k) = \frac{N}{e^{|\vec{\mathbf{k}}|/T}-1}+
\frac{N_f}{e^{|\vec{\mathbf{k}}|/T}+1}.
\eeqar{n_eq}
Here $N$ is the number of colors and $N_f$ is the number of flavors.

The matrix element ${\cal M}$ (see Eq.~(\ref{M_BT})), 
has been already computed in~\cite{BT_fermions} (see Eqs. (45-46) 
in~\cite{BT_fermions}). However, for completeness and due to a typographical 
error in~\cite{BT_fermions}, we here repeat the main steps in the computation 
of $\frac{1}{2} \sum_{spins} |{\cal M}|^2$.

\beqar
{\cal M}= g^2 
D_{\mu \nu } (q) \bar{u} (p^\prime, s^\prime) \gamma^\mu u (p, s) 
\bar{u} (k^\prime, \lambda^\prime) \gamma^\nu u (k, \lambda).
\eeqar{M_BT_App}

In Coulomb gauge, the only non-zero terms in the effective gluon propagator 
are given in Eqs.~(\ref{Pij}) and~(\ref{Q00}), which together with 
Eqs.~(\ref{dmnMed}-\ref{DeltaL}) reduce the Eq.~(\ref{M_BT_App}) to

\beqar
{\cal M} &=& g^2 \Delta_L (q) \bar{u} (p^\prime, s^\prime) \gamma^0 u (p, s) 
\bar{u} (k^\prime, \lambda^\prime) \gamma^0 u (k, \lambda) \nonumber \\
&+& g^2 \Delta_T (q) (\delta^{i j}-\hat{q}^i \hat{q}^j)
\bar{u} (p^\prime, s^\prime) \gamma^i u (p, s) 
\bar{u} (k^\prime, \lambda^\prime) \gamma^j u (k, \lambda).
\eeqar{calM1} 
Here $\hat{q}^i \equiv q^i/|\vec{\mathbf{q}}|$.

\medskip

The matrix element given in Eq.~(\ref{calM1}) has to be squared, averaged over 
initial spin $s$ of the jet and summed over all other spins. After evaluating 
the Dirac traces, and applying the assumption that $|\vec{\mathbf{q}}| \ll E$
(highly energetic jet) we obtain similarly to~\cite{BT_fermions} 

\beqar
\hspace*{3cm}\frac{1}{2} \sum_{spins} |{\cal M}|^2 &=& 16 \, g^4 \, E^2 \,
(|\Delta_L(q)|^2 (|\vec{\mathbf{k}}| |\vec{\mathbf{k^\prime}}|+
\vec{\mathbf{k}} \cdot \vec{\mathbf{k^\prime}})  \nonumber \\
&& \hspace*{-4.5cm} + \,2 Re(\Delta_L(q)\Delta_T(q)^*) 
\left[|\vec{\mathbf{k}}| (\vec{\mathbf{v}} \cdot \vec{\mathbf{k^\prime}} -
\frac{\vec{\mathbf{v}} \cdot \vec{\mathbf{q}} \; \vec{\mathbf{q}} \cdot \vec{\mathbf{k^\prime}}} {|\vec{\mathbf{q}}|^2})+
|\vec{\mathbf{k^\prime}}| (\vec{\mathbf{v}} \cdot \vec{\mathbf{k}} -
\frac{\vec{\mathbf{v}} \cdot \vec{\mathbf{q}} \; \vec{\mathbf{q}} \cdot \vec{\mathbf{k}}} {|\vec{\mathbf{q}}|^2})\right]\nonumber \\
&& \hspace*{-6cm}+ \,|\Delta_T(q)|^2 \left[2 \;(\vec{\mathbf{v}} \cdot \vec{\mathbf{k}} -
\frac{\vec{\mathbf{v}} \cdot \vec{\mathbf{q}} \; \vec{\mathbf{q}} \cdot \vec{\mathbf{k}}} {|\vec{\mathbf{q}}|^2})(\vec{\mathbf{v}} \cdot \vec{\mathbf{k^\prime}} -
\frac{\vec{\mathbf{v}} \cdot \vec{\mathbf{q}} \; \vec{\mathbf{q}} \cdot \vec{\mathbf{k^\prime}}} {|\vec{\mathbf{q}}|^2}) + 
(|\vec{\mathbf{k}}| |\vec{\mathbf{k^\prime}}|-
\vec{\mathbf{k}} \cdot \vec{\mathbf{k^\prime}}) 
(v^2- \frac{\vec{\mathbf{v}} \cdot \vec{\mathbf{q}} \; 
\vec{\mathbf{q}} \cdot \vec{\mathbf{v}}} {|\vec{\mathbf{q}}|^2}) \right] .
\eeqar{Cal_M_Sq}

In a static medium, the collisional energy loss does not depend on the 
direction of $\vec{\mathbf{v}}$. Therefore, the Eq.~(\ref{imm2a_App}) can be 
further simplified by averaging the integrand over the directions of 
$\vec{\mathbf{v}}$. The integrals that are required are

\beqar
{\cal J}_1 &=& \int \frac {d \Omega}{ 4 \pi } \; 
\frac{\sin[(\omega- \vec{\mathbf{v}}\cdot\vec{\mathbf{q}})\frac{L}{2 v}]^2}
{(\omega- \vec{\mathbf{v}}\cdot \vec{\mathbf{q}})^2} \; 
\nonumber \\
&=& \frac{L}{4 |\vec{\mathbf{q}}| v^2}
\left[Si( (v |\vec{\mathbf{q}}|+\omega)\frac{L}{v})+
Si( (v |\vec{\mathbf{q}}|-\omega)\frac{L}{v})\right] \nonumber \\ &-&
\frac{1}{4 v |\vec{\mathbf{q}}|} \left[\frac{1-\cos((v |\vec{\mathbf{q}}|-\omega)\frac{L}{v})}
{v |\vec{\mathbf{q}}|-\omega}+
\frac{1-\cos((v |\vec{\mathbf{q}}|+\omega)\frac{L}{v})}
{v |\vec{\mathbf{q}}|+\omega} \right] \; ,
\eeqar{J1}

\beqar
{\cal J}_2 &=& \int \frac {d \Omega}{ 4 \pi } \; 
\frac{\sin[(\omega- \vec{\mathbf{v}}\cdot\vec{\mathbf{q}})\frac{L}{2 v}]^2}
{(\omega- \vec{\mathbf{v}}\cdot \vec{\mathbf{q}})^2} \; (\omega- \vec{\mathbf{v}}\cdot \vec{\mathbf{q}})
\nonumber \\
&=& \frac{1}{4 v |\vec{\mathbf{q}}|} 
\left[Ci((v |\vec{\mathbf{q}}|-\omega)\frac{L }{v})-
Ci( (v |\vec{\mathbf{q}}|+\omega)\frac{L}{v})+
\ln(\frac{v |\vec{\mathbf{q}}|+\omega}{v |\vec{\mathbf{q}}|-\omega}) \right] 
\eeqar{J2}
and
\beqar
{\cal J}_3 = \int \frac {d \Omega}{ 4 \pi } \; 
\frac{\sin[(\omega- \vec{\mathbf{v}}\cdot\vec{\mathbf{q}})\frac{L}{2 v}]^2}
{(\omega- \vec{\mathbf{v}}\cdot \vec{\mathbf{q}})^2} \; (\omega- \vec{\mathbf{v}}\cdot \vec{\mathbf{q}})^2
=\frac{1}{2} \left(1-\frac{\cos(\frac{L \omega}{v}) \sin(L |\vec{\mathbf{q}}|)}
{L |\vec{\mathbf{q}}|} \right) \; .
\eeqar{J3}

By using Eqs.~(\ref{J1}-\ref{J3}), it can be shown that

\beqar
\int \frac {d \Omega}{ 4 \pi } \; 
\frac{\sin[(\omega- \vec{\mathbf{v}}\cdot\vec{\mathbf{q}})\frac{L}{2 v}]^2}
{(\omega- \vec{\mathbf{v}}\cdot \vec{\mathbf{q}})^2} \; v^i= 
{\cal J}_1 \hat{q}^i \frac{\omega}{|\vec{\mathbf{q}}|} - {\cal J}_2 
\frac{\hat{q}^i}{|\vec{\mathbf{q}}|} \; 
\eeqar{I2}
and
\beqar
\int \frac {d \Omega}{ 4 \pi } \; 
\frac{\sin[(\omega- \vec{\mathbf{v}}\cdot\vec{\mathbf{q}})\frac{L}{2 v}]^2}
{(\omega- \vec{\mathbf{v}}\cdot \vec{\mathbf{q}})^2} \; v^i v^j= 
{\cal J}_1 
\left( \frac{v^2 |\vec{\mathbf{q}}|^2-\omega^2}{|\vec{\mathbf{q}}|^2}
\delta^{ij}+\frac{3\omega^2- v^2 |\vec{\mathbf{q}}|^2}{ 2 |\vec{\mathbf{q}}|^2}
\hat{q}^i\hat{q}^j\right) + 
\frac{2 \omega {\cal J}_2 -{\cal J}_3}{2 |\vec{\mathbf{q}}|^2}
(\delta^{ij}-3 \hat{q}^i\hat{q}^j) \; .
\eeqar{I3}

By using Eqs.~(\ref{Cal_M_Sq}-\ref{I3}) it is straightforward to show that
\beqar
&&\int \frac {d \Omega}{ 4 \pi } \;  
\frac{\sin[(\omega- \vec{\mathbf{v}}\cdot\vec{\mathbf{q}})\frac{L}{2 v}]^2}
{(\omega- \vec{\mathbf{v}}\cdot \vec{\mathbf{q}})^2} \; 
16 \, g^4 \, E^2 \,
|\Delta_L(q)|^2 (|\vec{\mathbf{k}}| |\vec{\mathbf{k^\prime}}|+
\vec{\mathbf{k}} \cdot \vec{\mathbf{k^\prime}}) 
\nonumber \\
&& \hspace*{1.7cm} =16 \, g^4 \, E^2 \,
|\Delta_L(q)|^2 \frac{(2 |\vec{\mathbf{k}}|+\omega)^2- |\vec{\mathbf{q}}|^2}{2}
{\cal J}_1 \; ,
\eeqar{EL_1}

\beqar
&&\int \frac {d \Omega}{ 4 \pi } \;  
\frac{\sin[(\omega- \vec{\mathbf{v}}\cdot\vec{\mathbf{q}})\frac{L}{2 v}]^2}
{(\omega- \vec{\mathbf{v}}\cdot \vec{\mathbf{q}})^2} \; 
32 \, g^4 \, E^2 \, Re(\Delta_L(q)\Delta_T(q)^*) \nonumber \\
&& \hspace*{0.5cm} \times
\left[|\vec{\mathbf{k}}| (\vec{\mathbf{v}} \cdot \vec{\mathbf{k^\prime}} -
\frac{\vec{\mathbf{v}} \cdot \vec{\mathbf{q}} \; \vec{\mathbf{q}} \cdot 
\vec{\mathbf{k^\prime}}} {|\vec{\mathbf{q}}|^2})+
|\vec{\mathbf{k^\prime}}| (\vec{\mathbf{v}} \cdot \vec{\mathbf{k}} -
\frac{\vec{\mathbf{v}} \cdot \vec{\mathbf{q}} \; \vec{\mathbf{q}} \cdot 
\vec{\mathbf{k}}} {|\vec{\mathbf{q}}|^2})\right] =0
\eeqar{EL2}
and
\beqar
&&\int \frac {d \Omega}{ 4 \pi } \;  
\frac{\sin[(\omega- \vec{\mathbf{v}}\cdot\vec{\mathbf{q}})\frac{L}{2 v}]^2}
{(\omega- \vec{\mathbf{v}}\cdot \vec{\mathbf{q}})^2} \; 
16 \, g^4 \, E^2 \,
|\Delta_T(q)|^2 \nonumber \\
&& \hspace*{0.5cm} \times
\left[2 \;(\vec{\mathbf{v}} \cdot \vec{\mathbf{k}} -
\frac{\vec{\mathbf{v}} \cdot \vec{\mathbf{q}} \; \vec{\mathbf{q}} \cdot 
\vec{\mathbf{k}}} {|\vec{\mathbf{q}}|^2})(\vec{\mathbf{v}} \cdot 
\vec{\mathbf{k^\prime}} -
\frac{\vec{\mathbf{v}} \cdot \vec{\mathbf{q}} \; \vec{\mathbf{q}} \cdot 
\vec{\mathbf{k^\prime}}} {|\vec{\mathbf{q}}|^2}) + 
(|\vec{\mathbf{k}}| |\vec{\mathbf{k^\prime}}|-
\vec{\mathbf{k}} \cdot \vec{\mathbf{k^\prime}}) 
(v^2- \frac{\vec{\mathbf{v}} \cdot \vec{\mathbf{q}} \; 
\vec{\mathbf{q}} \cdot \vec{\mathbf{v}}} {|\vec{\mathbf{q}}|^2}) \right]  \,
|\Delta_T(q)|^2 \nonumber \\
&& \hspace*{0.5cm} =16 \, g^4 \, E^2 \,|\Delta_T(q)|^2 
\frac{(|\vec{\mathbf{q}}|^2-\omega^2)
((2 |\vec{\mathbf{k}}|+\omega)^2+ |\vec{\mathbf{q}}|^2)}
{4 |\vec{\mathbf{q}}|^4} \left[ (v^2 |\vec{\mathbf{q}}|^2-\omega^2) {\cal J}_1 
+2 \omega {\cal J}_2- {\cal J}_3 \right] \; .
\eeqar{EL3}

Therefore, averaging the 
$\frac{\sin[(\omega- \vec{\mathbf{v}}\cdot\vec{\mathbf{q}})\frac{L}{2 v}]^2}
{(\omega- \vec{\mathbf{v}}\cdot \vec{\mathbf{q}})^2} \; 
\frac{1}{2} \sum |{\cal M}|^2$ over the directions of 
$\vec{\mathbf{v}}$ lead to

\beqar
\left \langle 
\frac{\sin[(\omega- \vec{\mathbf{v}}\cdot\vec{\mathbf{q}})\frac{L}{2 v}]^2} 
{(\omega- \vec{\mathbf{v}}\cdot \vec{\mathbf{q}})^2} \; 
\frac{1}{2} \sum_{spins} |{\cal M}|^2 \right \rangle&=&
16 \, g^4 \, E^2 ( 
|\Delta_L(q)|^2 \frac{(2 |\vec{\mathbf{k}}|+\omega)^2  - 
|\vec{\mathbf{q}}|^2}{2} {\cal J}_1 \nonumber \\
&& \hspace*{-6.cm}
+ |\Delta_T(q)|^2 \frac{(|\vec{\mathbf{q}}|^2-\omega^2)
((2 |\vec{\mathbf{k}}|+\omega)^2+ |\vec{\mathbf{q}}|^2)}
{4 |\vec{\mathbf{q}}|^4} \left[(v^2 |\vec{\mathbf{q}}|^2-\omega^2) {\cal J}_1 +
2 \omega {\cal J}_2- {\cal J}_3   \right] )\; .
\eeqar{Average_el}

Since the collisional energy loss does not depend on the direction of 
$\vec{\mathbf{v}}$, the Eq.~(\ref{imm2a_App}) can be written as  

\beqar
\Delta E_{el} &=& C_R \frac{1}{E^2} \int \frac{d^{3} 
\vec{{\bf k}}}{(2 \pi)^{3} 2 k} 
n_{eq} (k)  \int \frac{d^{3} \vec{{\bf k^\prime}}}{(2 \pi)^{3} 2 k^\prime } \;  
\omega \; \left \langle 
\frac{\sin[(\omega- \vec{\mathbf{v}}\cdot\vec{\mathbf{q}})\frac{L}{2 v}]^2}
{(\omega- \vec{\mathbf{v}}\cdot \vec{\mathbf{q}})^2} \; 
\frac{1}{2} \sum_{spins} |{\cal M}|^2\right \rangle \nonumber \\
&=&  \frac{C_R}{32 \pi^4} \frac{1}{E^2} \int |\vec{\mathbf{k}}| 
|\vec{\mathbf{k}}^\prime| \, d |\vec{\mathbf{k}}| \, 
d |\vec{\mathbf{k}}^\prime| \, d\cos \theta \, n_{eq}(|\vec{\mathbf{k}}|)
\left \langle 
\frac{\sin[(\omega- \vec{\mathbf{v}}\cdot\vec{\mathbf{q}})\frac{L}{2 v}]^2}
{(\omega- \vec{\mathbf{v}}\cdot \vec{\mathbf{q}})^2} \; 
\frac{1}{2} \sum_{spins} |{\cal M}|^2\right \rangle \; ,
\eeqar{imm2a_App_average} 
where $\theta$ is the angle between vectors $\vec{\mathbf{k}}$ and 
$\vec{\mathbf{k}}^\prime$. Using the fact that $q=k^\prime-k$, we obtain that 
the $\cos \theta$ satisfies the following relation
\beqar
\cos \theta= 1-\frac{\vec{\mathbf{q}}^2- \omega^2}{2 |\vec{\mathbf{k}}|
|\vec{\mathbf{k}}^\prime|},
\eeqar{CosTheta}
where $|\vec{\mathbf{k}}^\prime|=|\vec{\mathbf{k}}|+\omega$. We can now 
introduce the change of variables from 
$|\vec{\mathbf{k}}|$, $|\vec{\mathbf{k}}^\prime|$ and $\cos \theta$, to 
$|\vec{\mathbf{k}}|$, $\omega$ and $|\vec{\mathbf{q}}|$, which reduces the 
Eq.~(\ref{imm2a_App}) to the following form 
\beqar
\Delta E_{el} &=&  \frac{C_R}{32 \pi^4} \frac{1}{E^2} 
\int n_{eq}(|\vec{\mathbf{k}}|) d |\vec{\mathbf{k}}| \; 
|\vec{\mathbf{q}}|d |\vec{\mathbf{q}}| \;
\omega d \omega \; \left \langle 
\frac{\sin[(\omega- \vec{\mathbf{v}}\cdot\vec{\mathbf{q}})\frac{L}{2 v}]^2}
{(\omega- \vec{\mathbf{v}}\cdot \vec{\mathbf{q}})^2} \; 
\frac{1}{2} \sum_{spins} |{\cal M}|^2\right \rangle \nonumber \\
&=& \frac{C_R g^4}{2 \pi^4}  
\int n_{eq}(|\vec{\mathbf{k}}|) d |\vec{\mathbf{k}}| \; 
|\vec{\mathbf{q}}|d |\vec{\mathbf{q}}| \; \omega d \omega \; 
( |\Delta_L(q)|^2 \frac{(2 |\vec{\mathbf{k}}|+\omega)^2  - 
|\vec{\mathbf{q}}|^2}{2} {\cal J}_1 \nonumber \\
&+& |\Delta_T(q)|^2 \frac{(|\vec{\mathbf{q}}|^2-\omega^2)
((2 |\vec{\mathbf{k}}|+\omega)^2+ |\vec{\mathbf{q}}|^2)}
{4 |\vec{\mathbf{q}}|^4} \left[ (v^2 |\vec{\mathbf{q}}|^2-\omega^2) {\cal J}_1 
+2 \omega {\cal J}_2- {\cal J}_3 \right] ) \; .
\eeqar{Average_3} 

Limits of the integration can be obtained from Eq.~(\ref{CosTheta}), from 
which it follows that
\beqar
0< \frac{\vec{\mathbf{q}}^2- \omega^2}{2 |\vec{\mathbf{k}}|
(|\vec{\mathbf{k}}|+\omega)}<2 \; ,
\eeqar{CosTheta_Limits}
leading to the limits in energy transfer $\omega$
\beqar
\text{Max}[{-|\vec{\mathbf{q}}| ,|\vec{\mathbf{q}}| 
-2 |\vec{\mathbf{k}}|}] <\omega< |\vec{\mathbf{q}}|.
\eeqar{omega_limits}

The limits on the momentum transfer $|\vec{\mathbf{q}}|$ from collisional scattering 
off a thermal parton with energy  $|\vec{\mathbf{k}}|$ is 
(see~\cite{TG})
\beqar
0 <|\vec{\mathbf{q}}|< \text{Min}[E, \frac{2 |\vec{\mathbf{k}}| 
(1-|\vec{\mathbf{k}}|/E)}{1-v+2|\vec{\mathbf{k}}|/E}].
\eeqar{q_limits}
Here $E$ and $v$ are the energy and velocity of the jet.

By using relations~(\ref{omega_limits}) and~(\ref{q_limits}), the 
Eq.~(\ref{Average_3}) finally reduces to
\beqar
\Delta E_{el} 
&=& \frac{C_R g^4}{2 \pi^4}  
\int_0^\infty n_{eq}(|\vec{\mathbf{k}}|) d |\vec{\mathbf{k}}| \; 
\left( \int_0^{|\vec{\mathbf{k}}|} |\vec{\mathbf{q}}|d |\vec{\mathbf{q}}| 
\int_{-|\vec{\mathbf{q}}|}^{|\vec{\mathbf{q}}|}\; \omega d \omega \;+
\int_{|\vec{\mathbf{k}}|}^{|\vec{\mathbf{q}}|_{max}} |\vec{\mathbf{q}}|d |\vec{\mathbf{q}}| 
\int_{|\vec{\mathbf{q}}|-2|\vec{\mathbf{k}}| }^{|\vec{\mathbf{q}}|}\; 
\omega d \omega \; \right)
 \nonumber \\
&& \hspace*{-2cm} \left( |\Delta_L(q)|^2 \frac{(2 |\vec{\mathbf{k}}|+\omega)^2  - 
|\vec{\mathbf{q}}|^2}{2} {\cal J}_1 +
|\Delta_T(q)|^2 \frac{(|\vec{\mathbf{q}}|^2-\omega^2)
((2 |\vec{\mathbf{k}}|+\omega)^2+ |\vec{\mathbf{q}}|^2)}
{4 |\vec{\mathbf{q}}|^4} (v^2 |\vec{\mathbf{q}}|^2-\omega^2) {\cal J}_1 +
2 \omega {\cal J}_2- {\cal J}_3 ) \right) ,\nonumber \\
\eeqar{Eel_result} 
where $|\vec{\mathbf{q}}|_{max}$ is given in Eq.~(\ref{q_limits}).

\subsection{Large $L$ limit}

In this subsection we will consider the large $L$ limit case, and compute the 
collisional energy loss per unit length. To do that we multiply both sides of 
the Eq.~(\ref{Average_3}) by $\frac{2 v}{\pi L}$ i.e.
\beqar
\frac{2 v}{\pi L} \Delta E_{el} &=&\frac{C_R g^4}{2 \pi^4}  
\int n_{eq}(|\vec{\mathbf{k}}|) d |\vec{\mathbf{k}}| \; 
|\vec{\mathbf{q}}|d |\vec{\mathbf{q}}| \; \omega d \omega \; 
( |\Delta_L(q)|^2 \frac{(2 |\vec{\mathbf{k}}|+\omega)^2  - 
|\vec{\mathbf{q}}|^2}{2} \frac{2 v}{\pi L} {\cal J}_1 \nonumber \\
&& \hspace*{-1.5cm} + |\Delta_T(q)|^2 \frac{(|\vec{\mathbf{q}}|^2-\omega^2)
((2 |\vec{\mathbf{k}}|+\omega)^2+ |\vec{\mathbf{q}}|^2)}
{4 |\vec{\mathbf{q}}|^4} (v^2 |\vec{\mathbf{q}}|^2-\omega^2) 
\frac{2 v}{\pi L} {\cal J}_1 +
2 \omega \frac{2 v}{\pi L}{\cal J}_2- \frac{2 v}{\pi L}{\cal J}_3 ) ).
\eeqar{Average3_limit} 
To compute $\frac{2 v}{\pi L}{\cal J}_{1, 2, 3}$ in the limit when 
$L\rightarrow \infty$ we will use the following expression
\beqar
\frac{2 v}{\pi L}
\frac{\sin[(\omega- \vec{\mathbf{v}}\cdot\vec{\mathbf{q}})\frac{L}{2 v}]^2}
{(\omega- \vec{\mathbf{v}}\cdot \vec{\mathbf{q}})^2} \; 
\xrightarrow{L\rightarrow \infty} 
\delta (\omega- \vec{\mathbf{v}}\cdot\vec{\mathbf{q}})\; .
\eeqar{delta_limit}

Then,
\beqar
\frac{2 v}{\pi L}{\cal J}_1 \;\xrightarrow{L\rightarrow \infty} 
\int \frac{d \Omega}{4 \pi} 
\delta (\omega- \vec{\mathbf{v}}\cdot\vec{\mathbf{q}}) = 
\frac{1}{2 v |\vec{\mathbf{q}}|} \Theta(v^2 |\vec{\mathbf{q}}|^2-\omega^2) \; ,
\eeqar{J1_limit}

\beqar
\frac{2 v}{\pi L}{\cal J}_2 \xrightarrow{L\rightarrow \infty} 
\int \frac{d \Omega}{4 \pi} \delta (\omega- \vec{\mathbf{v}}\cdot\vec{\mathbf{q}})
(\omega- \vec{\mathbf{v}}\cdot\vec{\mathbf{q}})= 0
\eeqar{J2_limit}
and

\beqar
\frac{2 v}{\pi L}{\cal J}_3 \xrightarrow{L\rightarrow \infty} 
\int \frac{d \Omega}{4 \pi} \delta (\omega- \vec{\mathbf{v}}\cdot\vec{\mathbf{q}})
(\omega- \vec{\mathbf{v}}\cdot\vec{\mathbf{q}})^2= 0 \; ,
\eeqar{J3_limit}

leading to
\beqar
\frac{2 v}{\pi L} \Delta E_{el} &=&\frac{C_R g^4}{2 \pi^4}  
\int n_{eq}(|\vec{\mathbf{k}}|) d |\vec{\mathbf{k}}| \; 
|\vec{\mathbf{q}}|d |\vec{\mathbf{q}}| \; \omega d \omega \; 
\frac{1}{2 v |\vec{\mathbf{q}}|} \Theta(v^2 |\vec{\mathbf{q}}|^2-\omega^2)
\nonumber \\
&& \hspace*{-2cm}\left( |\Delta_L(q)|^2 \frac{(2 |\vec{\mathbf{k}}|+\omega)^2  - 
|\vec{\mathbf{q}}|^2}{2}  + |\Delta_T(q)|^2 \frac{(|\vec{\mathbf{q}}|^2-\omega^2)
((2 |\vec{\mathbf{k}}|+\omega)^2+ |\vec{\mathbf{q}}|^2)}
{4 |\vec{\mathbf{q}}|^4} (v^2 |\vec{\mathbf{q}}|^2-\omega^2) \right).
\eeqar{Eel_limit} 

Therefore, by using relations~(\ref{omega_limits}),~(\ref{q_limits}) and 
$v^2 |\vec{\mathbf{q}}|^2>\omega^2$, the collisional energy loss per unit 
length in an infinite size QCD medium reduces the following expression ($C_R=4/3$)
\beqar
\frac{d E_{el}}{d L} &=&\frac{g^4}{6 \, v^2 \,\pi^3}  
\int_0^\infty n_{eq}(|\vec{\mathbf{k}}|) d |\vec{\mathbf{k}}| \; 
\left( \int_0^{|\vec{\mathbf{k}}|/(1+v)} d |\vec{\mathbf{q}}| 
\int_{-v |\vec{\mathbf{q}}|}^{v |\vec{\mathbf{q}}|}\; \omega d \omega \;+
\int_{|\vec{\mathbf{k}}|/(1+v)}^{|\vec{\mathbf{q}}|_{max}} d |\vec{\mathbf{q}}|
\int_{|\vec{\mathbf{q}}|-2|\vec{\mathbf{k}}| }^{v |\vec{\mathbf{q}}|}\; 
\omega d \omega \; \right)
\nonumber \\
&& \hspace*{-1cm}\left( |\Delta_L(q)|^2 \frac{(2 |\vec{\mathbf{k}}|+\omega)^2  - 
|\vec{\mathbf{q}}|^2}{2}  + |\Delta_T(q)|^2 \frac{(|\vec{\mathbf{q}}|^2-\omega^2)
((2 |\vec{\mathbf{k}}|+\omega)^2+ |\vec{\mathbf{q}}|^2)}
{4 |\vec{\mathbf{q}}|^4} (v^2 |\vec{\mathbf{q}}|^2-\omega^2) \right).
\eeqar{Eel_limit_Final} 

\end{appendix}

\end{document}